\documentclass[a4paper,onecolumn,11pt]{quantumarticle}
\pdfoutput=1
\usepackage[utf8]{inputenc}
\usepackage[english]{babel}
\usepackage[T1]{fontenc}
\usepackage{amsmath}
\usepackage{hyperref}

\usepackage{tikz}
\usetikzlibrary{arrows.meta}
\usepackage{quantikz}
\usetikzlibrary{decorations.pathreplacing}
\usepackage[margin=1.1in]{geometry}
\usepackage{amssymb,amsthm,mathtools}
\usepackage{booktabs}
\usepackage{enumitem}
\usepackage[expansion=false]{microtype}
\usepackage{comment}
\usepackage{graphicx}
\usepackage{subcaption}
\theoremstyle{plain}
\newtheorem{theorem}{Theorem}[section]
\newtheorem{proposition}[theorem]{Proposition}
\newtheorem{corollary}[theorem]{Corollary}
\newtheorem{lemma}[theorem]{Lemma}
\theoremstyle{definition}
\newtheorem{definition}[theorem]{Definition}
\newtheorem{remark}[theorem]{Remark}

\newcommand{\R}{\mathbb{R}}
\newcommand{\C}{\mathbb{C}}
\newcommand{\T}{\mathbb{T}}
\newcommand{\Cay}{\mathcal{C}}
\newcommand{\Rat}{\mathcal{R}}
\newcommand{\Lam}{\Lambda}
\newcommand{\norm}[1]{\left\lVert #1 \right\rVert}
\newcommand{\Ot}{\widetilde{O}}
\DeclareMathOperator{\arccot}{arccot}

\DeclareMathOperator{\sign}{sign}
\usepackage{pgfplots}
\pgfplotsset{ignore legend/.style={every axis legend/.code={\let\addlegendentry\relax}}}
\pgfplotsset{compat=1.18}
\begin{document}
\usepgfplotslibrary{groupplots}

\title{Cayley access models for matrix functions: reachable class and cost}
\author{Gianna M. Del Corso}
\affiliation{Department of Computer Science, University of Pisa, Italy }
\orcid{0000-0002-5651-9368}
\email{gianna.delcorso@unipi.it}
\date{}

\maketitle
\begin{abstract}
We study how to compute $f(A)b$ on a quantum computer, where $A$ is Hermitian.  When $f$ is nonsmooth or has localized difficulty in the spectrum, classical numerical linear algebra approximates it with rational functions rather than polynomials.

We propose the Cayley transform, parameterized by $\sigma \in \mathbb{R}$ and $\gamma > 0$, as an alternative  to the standard quantum access model. The two parameters let one concentrate accuracy where $f$ is difficult, which standard models can achieve only by raising the polynomial degree globally. This transformation converts Hermitian matrices to unitaries, allowing generalized quantum signal processing to apply Laurent polynomials bounded on the unit circle. 

We compare three access models (block encoding, matrix exponential, and Cayley encoding) in terms of query complexity and of the cost of their circuit realization. When the difficulty of $f$ is localized in the spectrum, the Cayley encoding reduces the required degree by a factor of three to four, and it represents the shifted resolvent exactly at degree one. For spectral projectors, the standard models cannot reach the target accuracy at fixed degree, because their approximants cannot concentrate resolution near the discontinuity.

Realizing the Cayley encoding from a block encoding requires solving a shifted linear system, whose cost offsets its query-complexity gain over the standard models. For structured matrices arising in practice, however, low-depth circuits for the encoding can be constructed directly, and the gain is preserved at the circuit level.

\end{abstract}

\section{Introduction}

Many computational tasks reduce to applying a function of a matrix to a vector.
Given a Hermitian $A\in\C^{N\times N}$ with eigenpairs $Av_j=\lambda_jv_j$, a
scalar function $f$ acts on the eigenvalues as
\begin{equation}\label{eq:fdef}
f(A) = \sum_{j} f(\lambda_j)\, v_jv_j^*,
\end{equation}
and the problem we consider in this paper is to compute $f(A)\,b$ for a given $b$. Different choices of
$f$ give rise to a wide range of tasks: $f(\lambda)=1/\lambda$ solves the linear
system $Ax=b$; a smoothed indicator of an interval gives the spectral
projector, which filters $b$ onto the eigenvectors with eigenvalue in that
interval and whose trace counts them; the sign function separates the positive
and negative parts of the spectrum;
$e^{-i \lambda t}$, $\lambda^{-1/2}$ occur in dynamics, sampling and
preconditioning. For matrices of moderate size several methods are
available~\cite{Hig08}, but when $N$ is very large, forming or diagonalizing $A$
is out of reach and one looks for algorithms whose cost is polynomial in
$\log N$.  This is the setting of quantum linear algebra, and the one considered in this
paper.
%

Computing $f(A)b$ on a quantum computer poses two distinct subproblems.
\begin{itemize}
\item {Approximating $f$.} A quantum computer applies  unitary operations and
reaches $A$ only through combinations of a supplied operator, so it cannot
evaluate an arbitrary $f$: it applies to the eigenvalues of $A$ an approximant
of $f$ rather than $f$ itself,  as done in  classical computation. That approximant is usually a polynomial, and its degree
controls the cost. When $f$ is nonsmooth, or has a discontinuity or a pole near the spectrum,
polynomials converge slowly and the degree becomes prohibitive. Classical
numerical linear algebra turns to rational approximation in  these
cases, and we ask here whether the same can be done on a quantum computer.
\item{Encoding $A$.} A quantum  algorithm cannot see $A$ directly; it accesses it through  an {access model}, that is a unitary circuit, from which $A$ can be inferred. The reachable functions are dictated jointly by the access model and by the transformation applied to it. Two access models are standard, block encoding~\cite{GSLW19} and matrix-exponential (also known as Hamiltonian simulation)~\cite{LC17}.
\end{itemize}
In this paper we investigate the \emph{Cayley transform} as an alternative access model for a Hermitian $A$ when the goal is to compute $f(A)b$. It is the matrix version of the M\"obius map $z_{\sigma,\gamma}(\lambda)=(\lambda-\sigma-i\gamma)/(\lambda-\sigma+i\gamma)$ sending the real line to the unit circle, and applied to $A$ it produces a unitary matrix. The logic of the construction is the following; GQSP~\cite{MW24} already applies to a unitary any Laurent polynomial of it
bounded by one on the unit circle, with one ancilla qubit. Taking $\Cay_{\sigma,\gamma}(A)$ as that unitary, whose eigenvalues are M\"obius images of the eigenvalues of $A$, the Laurent polynomial becomes, back in $\lambda$, a {rational} function of $A$. 

We concentrate our study on the following aspects. We show that the reachable functions are the rational functions of type $(2d,2d)$ whose poles lie at $\sigma\pm i\gamma$, each of order at most $d$.
Unlike polynomials, this class represents exactly functions with poles at $\sigma\pm i\gamma$, but it does not coincide with the rational functions of the same degree with freely chosen poles.
Multiplying Cayley factors extends it to arbitrary pole configurations, and we show that the resulting family is the only one available, since the finite products of Cayley factors are the rational functions unitary on every Hermitian matrix.

Our model has tunable parameters that adapt the encoding to the problem at
hand. In the standard access models the unitary is fixed by $A$ alone, the
normalization $\alpha$ and the evolution time $t$ depending only on the matrix,
and only the polynomial is constructed on $f$. Here $\sigma$ and $\gamma$ decide
where in the spectrum the approximation is accurate and how finely, so choosing
them already records where $f$ is difficult.  As in the matrix-exponential model, one ancilla suffices and no
subnormalization constant appears.

The cost of the construction depends on how the encoding is realized. In the
query model, where $\Cay_{\sigma,\gamma}(A)$ is supplied as an oracle, the
degree needed is most of the time far lower than for the two standard models.
Building that oracle from a block encoding of $A$ requires solving a shifted
linear system, whose cost has to be weighed against the saving. On a generic matrix, this cost per query cancels the reduction in the number of queries, so the total cost is of the same order as in the standard models. The advantage therefore survives only for matrices whose structure allows the Cayley encoding to be built directly as a low-depth circuit.
\subsection{Contributions}\label{sec:contrib}

This paper proposes the Cayley transform as a rational access model for matrix
functions on a quantum computer, and examines what it can compute, what it
costs, and when it is worth using.
At the level of {expressive power}, we show that processing $\Cay_{\sigma,\gamma}(A)$ by GQSP with a Laurent polynomial of
degree $d$ reaches the rational functions of type $(2d,2d)$ whose poles lie
at $\sigma\pm i\gamma$, each of order at most $d$, with one ancilla and no
parity restriction (Theorem~\ref{thm:class}). Approximating $f$ then reduces to trigonometric
approximation on the circle (Proposition~\ref{prop:reduction}), so the
classical theory of polynomial approximation carries over to a rational problem
through a change of variable. The parameters $\sigma$ and $\gamma$ enter the
encoding itself, concentrating the accuracy where  $f$ is difficult before any
polynomial is chosen, which the fixed access models cannot do. The three access models become comparable through the accuracy density, the
derivative of the map each of them induces on the spectrum, which measures how
much phase a model assigns to a region of the spectrum and therefore how finely a given degree can distinguish eigenvalues there.

Products of Cayley factors place the poles at any finite set of points off the
real axis, still with a single ancilla. We show that these are exactly the
rational functions unitary on every Hermitian matrix
(Theorem~\ref{thm:completeness}), so no other rational unitary encoding
exists, and that on the targets considered here the extra poles bring no
advantage.

On the side of cost we separate the query model from the realization model. In
the query model, where the encoding is supplied as an oracle, the number of
applications is smaller than for the two standard models on the functions that
polynomials approximate slowly. Whether this carries over to circuit cost
depends on how the encoding is built. In Section~\ref{sec:cost} we realize
$\Cay_{\sigma,\gamma}(A)$ from a block encoding of $A$, and we show that the
reduction in degree that a narrow $\gamma$ buys is spent, at the same rate, in
the shifted solve that each application then requires: the two parameters do not
by themselves lower the cost. What the analysis leaves is a condition on $A$, namely that the encoding admit a
circuit whose cost does not grow with $1/\gamma$, and in
Section~\ref{sec:structured} we identify classes of matrices of practical
interest that meet it.
Finally, we validate the analysis numerically (Section~\ref{sec:numerics}) by
computing the best approximation in each reachable class, with the free
parameter of each model optimized at every degree. The comparison is one of
approximating power and of query counts; no circuit is simulated and no gate
count is reported. At accuracy $10^{-6}$ the Cayley class is the cheapest on six
out of eight functions. The matrix exponential usually lies between the other
two. It, too, is applied by processing a unitary, namely $e^{-iAt}$, so the fact
that the Cayley encoding needs fewer queries cannot be attributed to unitarity by
itself: the gain instead comes from placing poles near a difficulty of $f$,
which rational functions approximate better than polynomials can. On the family of Sturm--Liouville operators of
Section~\ref{sec:numerics-jacobi} we measure the quantities that govern the
structured realization. For functions whose variation is concentrated near the
smallest eigenvalue, the Cayley class needs a degree independent of $N$, while
the polynomial and the exponential approaches need a degree of order $\kappa$;
with the encoding realized from the generators of the shifted resolvent, this
gives a cost of $\Ot(\kappa^{1/4})$ against $\Ot(\kappa)$.

\subsection{Relation to prior work}\label{sec:related}

Classically, there are many tools one can use to compute $f(A)b$:
shift-and-invert and rational Krylov methods~\cite{Gut13}, contour-integral
eigensolvers~\cite{SS03}, Zolotarev's optimal rational approximants~\cite{NF16},
complex-shifted preconditioners for indefinite systems~\cite{EVO04}, and the
Crank--Nicolson propagator~\cite{HLW06}. Which of them performs best depends on
the spectrum of $A$ and on the regularity of $f$, and rational approximations
have become the standard tool when polynomial ones converge slowly, as happens
with singularities close to the spectrum or poorly separated eigenvalues.

The quantum setting has followed a similar path. Early algorithms relied on
quantum phase estimation, which gives coherent access to the spectral
decomposition: the Harrow--Hassidim--Lloyd algorithm implements
$f(\lambda)=1/\lambda$, and later work extended the idea to fractional powers,
exponentials, Gibbs states~\cite{PW09} and linear differential
equations~\cite{BerryChildsOstranderWang2017}. These methods need
high-precision phase estimation and long coherent evolutions, which motivated
frameworks acting directly on the spectrum. Quantum signal
processing~\cite{LC17} showed that polynomial transformations of the
eigenvalues of a unitary can be implemented by sequences of single-qubit
rotations, and the quantum singular value transformation~\cite{GSLW19,MRTC21}
extended this to the singular values of a block-encoded matrix, unifying
Hamiltonian simulation, linear system solvers, matrix inversion and
Gibbs-state preparation. Building a block encoding is itself costly, and for a sparse matrix it needs several ancillas and multiqubit controlled operations. This problem was solved by two later frameworks, which take as input the time
evolution $e^{-iA}$,  produced directly by a Trotter formula, and use a single
ancilla. The first is quantum eigenvalue transformation of unitary
matrices~\cite{DLT22}, which needs no multiqubit control either: writing
$e^{i\Phi}$ for an eigenvalue of the unitary, the method reaches the functions
$F(\cos(\Phi/2))$ with $F$ an even real polynomial. 
The second is the generalized version of signal processing~\cite{MW24,Su23},
which requires controlled access to the unitary but allows general $SU(2)$
rotations, and reaches every Laurent polynomial bounded by one on the circle,
with complex coefficients and no parity restriction.

All of these frameworks are polynomial, and a polynomial is the wrong tool when
the best approximant of $f$ is rational. Two lines of work have addressed this.
Contour-integral algorithms start from Cauchy's formula and apply a quadrature
rule to the integral, so that $f(A)b$ is obtained from a set of shifted linear
systems $(z_kI-A)x_k=b$ weighted by the quadrature
coefficients~\cite{Takahira20}. Shen, Van Buggenhout, Camps, Klymko and Van
Beeumen~\cite{SVCKV24} instead implement $r(A)=\sum_k c_k(z_kI-A)^{-1}$
directly, with poles chosen at will and each resolvent built from linear
combinations of matrix exponentials, at the price of $\lceil\log_2 m\rceil$
ancillas and a subnormalization factor for the combination.

Our starting point is different. Rather than combining several resolvents, the
Cayley transform turns the computation of a matrix function into a unitary
signal-processing problem, with one ancilla and no overhead from a linear
combination. It allows arbitrary pole placement, through products of Cayley
factors, but the numerator is not free, being tied to the poles by the
unitarity of the encoding. The construction therefore lies between the
polynomial signal-processing methods and the rational approximations with a
free numerator, and this intermediate position is a structural necessity. By Theorem~\ref{thm:completeness} the class of rational functions unitary on every Hermitian matrix coincides with the finite products of Cayley factors, so no general rational encoding can be unitary and a method with a free numerator necessarily leaves the single-ancilla setting.

Quantum Krylov methods, on the other hand, are still polynomial: the subspace is
generated from real-time evolutions~\cite{Klymko2022,Kirby2023} and not from
resolvents, so their classical rational counterpart has no quantum analogue
yet.

Tong, An, Wiebe and Lin~\cite{TAWL21} consider a normal $A$ with an efficiently
implementable diagonalizing unitary, and observe that a block encoding of
$A^{-1}$ can then be built directly, by writing the eigenvalue in a register and
inverting it with classical arithmetic, at a cost independent of $\kappa(A)$.
They use it as a preconditioner for $(A+B)x=b$ with $\norm A\gg\norm B$, and to
evaluate $e^{-\beta (A+B)}$ through a quadrature of Cauchy's formula, that is again
through a linear combination of resolvents. The block encoding of $A$ is thereby
never formed, which is also what happens in Section~\ref{sec:structured}, under
a different structural hypothesis: fast
inversion needs a circuit for the eigenvector matrix of $A$ and a formula for its
eigenvalues, whereas our construction needs the inverse of the shifted matrix to
be one-pair semiseparable, as for every irreducible tridiagonal $A$, and its
generators to be prepared by a low-depth circuit. 

\section{Preliminary Results and Notation}\label{sec:notation}
Throughout the paper $A\in\C^{N\times N}$ is Hermitian, so that $A=\sum_j\lambda_jv_jv_j^*$ with $\lambda_j\in\R$, $\{v_j\}$ orthonormal. Denote by $\Lam(A)$ the spectrum, $\norm{\cdot}$ the spectral norm, $\kappa(\cdot)$ the condition number, $\T$ the unit circle. Since $\Lam(A)$ is not known in practice, functions of $A$ are approximated on
a known closed set $S\supseteq\Lam(A)$, which we call the spectral set; in the
experiments it is a finite union of closed intervals. For $w\notin\Lam(A)$, $R_w(A)=(A-wI)^{-1}$ is the resolvent, and $R_w(A)b$ solves the shifted system $(A-wI)x=b$. We use one fact. 
For $w=\sigma+i\tau$ with $\tau\neq0$ the matrix $A-wI$ is normal, being a
polynomial in $A$, with eigenvalues $(\lambda_j-\sigma)-i\tau$ of modulus
between $|\tau|$ and $\sqrt{\|A-\sigma I\|^{2}+\tau^{2}}$; since for a normal
matrix the singular values are these moduli,
\begin{equation}\label{eq:cond}
\kappa(A-wI)\le\frac{\sqrt{\|A-\sigma I\|^{2}+\tau^{2}}}{|\tau|}
\approx\frac{\|A-\sigma I\|}{|\tau|}\qquad( \mbox{ if }|\tau|\ll\|A-\sigma I\|),
\end{equation}
independently of any small eigenvalue of $A$: the shifted system is well
conditioned even when $A$ is singular.

This paper is built around the Cayley encoding of $A$.
\begin{definition}\label{def:cayley}
For $A=A^*$, $\sigma\in\R$, $\gamma>0$,\footnote{For $\gamma=0$ the encoding reduces
to the identity, and the sign of $\gamma$ is a normalization, since
$\Cay_{\sigma,-\gamma}(A)=\Cay_{\sigma,\gamma}(A)^{*}$.} the Cayley encoding of $A$ with centre $\sigma$ and width $\gamma$ is
\[
\Cay_{\sigma,\gamma}(A)=(A-\sigma I-i\gamma I)(A-\sigma I+i\gamma I)^{-1}.
\]
Its inverse, defined for a unitary $U$ with $1\notin\Lam(U)$, is $\Cay^{-1}_{\sigma,\gamma}(U)=\sigma I+i\gamma(I+U)(I-U)^{-1}$; the two are mutually inverse bijections between Hermitian matrices and unitaries without the eigenvalue $1$.
\end{definition}

\begin{proposition}\label{prop:cayley}
$\Cay_{\sigma,\gamma}(A)$ is unitary, commutes with $A$, and shares its eigenvectors; its eigenvalues are $z_{\sigma,\gamma}(\lambda_j)=(\lambda_j-\sigma-i\gamma)/(\lambda_j-\sigma+i\gamma)\in\T\setminus\{1\}$.
\end{proposition}

The next identity relates the encoding to the resolvent of $A$, hence to the shift-and-invert operator of classical numerical linear algebra. It is purely algebraic, independent of how the encoding is realized on a quantum computer.

\begin{proposition}[resolvent form]\label{prop:resolvent}
$\Cay_{\sigma,\gamma}(A)=I-2i\gamma\,R_{\sigma-i\gamma}(A)$.
\end{proposition}


In Section~\ref{sec:cost} we describe how to realize 
$\Cay_{\sigma, \gamma}(A)$ on a quantum computer and the cost of this realization. As an access model, it is treated as an oracle, similarly to how $e^{-iAt}$ is in the matrix-exponential model.

\noindent \textbf{ Rational functions}. A rational function $r=p/q$, with $p,q$
coprime, has a pole at $w$ when $q(w)=0$, of order the multiplicity of that
root; it has type $(\mu,\nu)$ if $\deg p\le\mu$ and $\deg q\le\nu$, and then at
most $\nu$ poles counted with order. For a finite set $\mathcal{P}\subset\C$ we write
$\Rat_{\mu,\nu}(\mathcal{P})$ for the type-$(\mu,\nu)$ rational functions with poles
in $\mathcal{P}$.

For $\gamma>0$ we write
$\Rat_d(\sigma,\gamma)=\{\,N(\lambda)/((\lambda-\sigma)^2+\gamma^2)^{d}:\ \deg N\le 2d\,\}$,
the rational functions of type $(2d,2d)$ whose poles lie at $\sigma\pm i\gamma$,
each of order at most $d$; it is a linear space of dimension $2d+1$, and
$\Rat_d(\sigma,\gamma)\subset\Rat_{2d,2d}(\{\sigma\pm i\gamma\})$, strictly for
$d\ge1$, since the latter also contains functions with a single pole of order
$2d$, such as $(\lambda-\sigma+i\gamma)^{-2d}$.

\subsection{Signal processing}\label{sec:gqsp}
In this section we describe the  generalized quantum signal processing (GQSP)~\cite{MW24}, the transformation mechanism used with the Cayley encoding and the matrix-exponential model.
 Given controlled access to a unitary $U$ and to $U^*$, one interleaves them with single-qubit $SU(2)$ rotations $R_k$ (three Euler angles each) on one ancilla qubit; the alternating circuit realizes, in the $\bra{0}\cdot\ket{0}$ block,
\[
P(U)=\sum_{k=-d}^{d}c_k\,U^k,
\]
an arbitrary Laurent polynomial subject only to $|P(z)|\le1$ on $\T$. When $U$ and $A$ share eigenvectors, if $Uv=e^{i\varphi}v$ then $P(U)v=P(e^{i\varphi})v$: a Laurent polynomial in $U$ applies the scalar map $e^{i\varphi}\mapsto P(e^{i\varphi})$ to each eigenphase. Two features distinguish GQSP from earlier QSP: general $SU(2)$ rotations remove the parity and reality restrictions and the need for a complementary polynomial, leaving only boundedness on $\T$; and the angles are computed by a stable classical routine depending only on $P$, and hence on the target function $f$, and not on $A$ or the dimension $N$, a portable precomputation obtained by layer-stripping recursions~\cite{MW24} or by direct optimization~\cite{DMWL21}.

Choosing an access model is thus choosing which unitary $U$ carries the eigenvalues of $A$, and with it the classical approximation problem. This paper takes $U=\Cay_{\sigma,\gamma}(A)$, whose eigenphases are M\"obius
images of $\lambda$, so that $P(U)$ is a rational function of $A$; the two
standard choices, which yield polynomials, are compared with it in
Section~\ref{sec:threemodels}.

\section{The Cayley encoding} \label{sec:cayley}

Since $\Cay_{\sigma,\gamma}(A)$ is unitary, it
can be used as an access model and processed by GQSP. What makes it useful is that it carries two free
parameters. This section studies the effect of $\sigma$, $\gamma$ and the degree $d$. The
derivative of the eigenphase map decides where in the spectrum, and how
finely, the encoding acts, and hence how well a given $f$ can be approximated.

\textbf{Example.} A reference example that we will develop in the paper is the shifted resolvent
$R_w(A)$, $w=\sigma-i\gamma$: this is what shift-and-invert and rational Krylov methods perform, and the Cayley encoding realizes it at degree one (Section~\ref{sec:resolvent}).

\subsection{The two parameters}\label{sec:params}

By Proposition~\ref{prop:cayley} each eigenvalue $\lambda$ of $A$ is carried to
an eigenphase $e^{i\varphi_{\sigma,\gamma}(\lambda)}$ of the encoding, with
\begin{equation}\label{eq:phasemap}
\varphi_{\sigma,\gamma}(\lambda)=-2\arccot\!\Bigl(\tfrac{\lambda-\sigma}{\gamma}\Bigr),\qquad
\lambda(\varphi)=\sigma-\gamma\cot(\varphi_{\sigma,\gamma}/2),
\end{equation}
a strictly increasing bijection from $\R$ onto $(-2\pi,0)$: the whole circle is
covered except the point $z=1$, the common image of $\lambda\to\pm\infty$. Distinct eigenvalues therefore receive distinct
phases for every $\gamma>0$, so unlike $e^{-iAt}$ there is no aliasing and no
need to know $\norm A$. In the following, we write $\varphi$ for $\varphi_{\sigma, \gamma}$
 when the parameters are fixed

The value assigned to $\lambda$ is $P(e^{i\varphi(\lambda)})$, where $P$ is a
Laurent polynomial of degree $d$, that is a trigonometric polynomial of degree
$d$ in the angle $\varphi$. If two eigenvalues $\lambda_1,\lambda_2$ are close
but $f(\lambda_1)$ and $f(\lambda_2)$ differ by an amount of order one, as
happens across a jump or a steep region of $f$, then any approximant with
small error must take values differing by about the same amount at
$\lambda_1$ and $\lambda_2$. By Bernstein's inequality~\cite{DL93} a
trigonometric polynomial bounded by one cannot change by an amount of order
one over an interval of angles shorter than about $1/d$. Since
$\varphi(\lambda_2)-\varphi(\lambda_1)=\varphi'(\hat\lambda)(\lambda_2-\lambda_1)$
for some $\hat\lambda$ between $\lambda_1$ and $\lambda_2$, two eigenvalues can
therefore receive substantially different values only if
\begin{equation}\label{eq:resolution}
|\lambda_1-\lambda_2|\;\gtrsim\;\delta(\hat\lambda),
\qquad
\delta(\lambda):=\frac{1}{d\,\varphi'(\lambda)},
\qquad
\varphi'(\lambda)=\frac{2\gamma}{(\lambda-\sigma)^2+\gamma^2}.
\end{equation}
Two eigenvalues closer than $\delta(\hat\lambda)$ are assigned almost the same
value, whatever the polynomial of degree $d$. The estimate is of order of
magnitude only, since Bernstein's inequality bounds the derivative of $P$ on
the whole circle and not on the arc between $\varphi(\lambda_1)$ and
$\varphi(\lambda_2)$.

The function $\varphi'$ is a bell-shaped weight with $\int_\R\varphi'=2\pi$, peaked at
$\sigma$ with value $2/\gamma$ and decaying as $2\gamma/(\lambda-\sigma)^2$, so
$\delta(\sigma)$ equals $\gamma/(2d)$, stays of that order on a band
of width about $\gamma$ around it, and grows quadratically outside. The center
$\sigma$ thus is where the encoding acts finely, the width $\gamma$ over
how large a region, and the degree $d$ improves the resolution by the same
factor everywhere.

Let $\Delta$ be the scale on which $f$ varies rapidly: the width of a jump or
of a steep region of $f$, or the distance from $S$ of a singularity of $f$
lying close to it.

Since $\delta(\sigma)=\gamma/(2d)$ by~\eqref{eq:resolution}, distinguishing two
eigenvalues $\Delta$ apart near $\sigma$ requires $\gamma/(2d)\lesssim\Delta$. Bringing the error below $\varepsilon$ requires in addition a degree growing
like $\log(1/\varepsilon)$, by the geometric decay of trigonometric
approximation for $f$ analytic near $S$~\cite{Tr19}. Combining the two requirements,
\begin{equation}\label{eq:ddouble}
d\ \gtrsim\ \max\Bigl\{\ \frac{\gamma}{2\Delta}\ ,\ c\,\log\frac1\varepsilon\ \Bigr\},
\end{equation}
with $c$ a constant depending on $f$. The first term grows with $\gamma$ and
the second does not, which makes the choice of $\gamma$ an optimization rather
than a change of variables; Section~\ref{sec:budget} works out the trade-off
once the cost of one application is included.

\textbf{Example} For the shifted resolvent at $\sigma-i\gamma$ the two
parameters are obtained from the function itself: $\sigma$ is the real part of the
evaluation point and $\gamma$ its distance from the real axis. The band of
width about $\gamma$ around $\sigma$, where $\varphi'$ is largest, then lies where the resolvent varies fastest, and the resolvent is obtained
  at degree one in closed form in Section~\ref{sec:resolvent}. The smaller $\gamma$, the closer the pole to the spectrum and the harder the
function is for a polynomial method, which has no pole to place; the encoding
instead pays for a small $\gamma$ in the cost of one application, which grows
as $1/\gamma$ (Section~\ref{sec:cost}).

\textbf{Choosing the parameters in practice.} The parameters are set by $f$ and
not by the eigenvalues of $A$, which are usually unknown: $\sigma$ goes where
$f$ needs accuracy and $\gamma$ is the scale of the region where the closest
eigenvalues must be assigned different values. For a window, or for eigenvalues
near a point $\tau$, one takes $\sigma=\tau$ and $\gamma$ about the half-width.
The choice is a starting point rather than an optimum, since the cost of a
degree-$d$ approximant of width $\gamma$ also depends on $\gamma$ through the
cost of one application; Section~\ref{sec:numerics} shows that the pair minimizing the effective cost is
not the one of smallest degree, a wider $\gamma$ being often preferable because
it lowers the rescaling.

\subsection{Computing \texorpdfstring{$f(A)$}{f(A)}} \label{sec:fA}
By Proposition~\ref{prop:cayley}, $U=\Cay_{\sigma,\gamma}(A)$ can serve as a
unitary encoding. We now use it as the GQSP signal to reach $f$. A single
application of GQSP with a Laurent polynomial $P$ of degree $d$ produces
$r(A)=P(U)$, with $r\in\Rat_d(\sigma,\gamma)$, and since $A$ is Hermitian and
$U$ commutes with it,
\[
\norm{f(A)-P(U)}=\max_{\lambda\in\Lam(A)}\bigl|f(\lambda)-P(z_{\sigma,\gamma}(\lambda))\bigr| ,
\]
so choosing $P$ is a scalar approximation problem on the spectrum. Recovering
$r(A)b$ from the $\ket0$ branch of the ancilla requires amplitude amplification
(step 6 below).

We now assemble the full procedure, analyzed and discussed in the subsequent sections. Its overall flow is
\[
\boxed{f}\ \xrightarrow{\text{choose } \sigma, \gamma}\ \boxed{F \text{ on } \T}\ \xrightarrow{\text{approximate}}\ \boxed{P,\ \{c_k \}}\ \xrightarrow{\text{angle-finding}}\ \boxed{\{R_k\}}\ \xrightarrow[ U=\Cay_{\sigma,\gamma}(A)]{\text{GQSP}}\ \boxed{r(A)\,b},
\]
where the first four stages are classical precomputation (independent of $N$, and of $A$ from the angle-finding on) and only the last runs on the quantum device. In detail:
\begin{enumerate}[leftmargin=*]
\item Choose  $\sigma$ where the target  function needs accuracy, $\gamma$ the scale of the sharp region, as discussed at the end of Section~\ref{sec:params}.

\item Set $F(\varphi)=f(\sigma-\gamma\cot(\varphi/2))$. This $F$ is continuous whenever $f$ is continuous on the spectrum.
\item  Approximate $F$ by a degree-$d$ trigonometric polynomial $p$ (Section~\ref{sec:approx}) and let $\eta$ be the supremum of $p$ on $\T$.
Dividing the coefficients of $p$ by $\eta$ gives $P(e^{i\varphi})=\sum_k c_k
e^{ik\varphi}$ with $|P|\le1$, as GQSP requires. Note that the constraint acts on the whole circle, whereas $f$ prescribes $p$ only on the
arc $z_{\sigma,\gamma}(S)$, the image of the spectral set; off that arc nothing
holds $p$ down, so $\eta$ can exceed the maximum of $f$ and is not a property of
$f$ alone but of how much of the circle the spectral set covers. In Proposition~\ref{prop:reduction} we show that
this is equivalent to prescribed-pole rational approximation of $f$, but one
works only with the polynomial.
\item Compute the GQSP angles from $\{c_k\}$ obtaining the rotations $\{ R_k\}$; a matrix-inde\-pen\-dent, reusable precomputation.
\item Build the circuit: $2d+1$ single-qubit rotations $\{ R_k\}$ interleaved with $2d$ controlled applications of $\Cay_{\sigma,\gamma}(A)^{\pm1}$ (Section~\ref{sec:cost} discusses how the encoding is realized).
\item Run on $\ket0\ket b$ and measure the ancilla. On outcome $0$ the system register holds $P(\Cay_{\sigma,\gamma}(A))\,b$, which approximates $\eta^{-1}f(A)b$, where $\eta$ is the factor introduced in step~3. The outcome has probability $\norm{P(\Cay_{\sigma,\gamma}(A))b}^2$, of order $\bigl(\norm{f(A)b}/\eta\bigr)^2$, so the vector is extracted by amplitude amplification, at a cost quantified in Corollary~\ref{cor:endtoend-vector}.
\end{enumerate}


\subsection{The class reachable with a Cayley factor}\label{sec:reachable}

In this section we describe the main steps for the approximation of $f(A)b$ using Cayley encoding and a Laurent polynomial of degree $d$.

\begin{theorem}\label{thm:class}
For a Laurent polynomial $P$ of degree $d$, $P(\Cay_{\sigma,\gamma}(A))=r(A)$ with $r\in\Rat_d(\sigma,\gamma)$. for every $r\in\Rat_d(\sigma,\gamma)$ bounded
by $1$ on $\R$ there exists a unique Laurent polynomial $P$ of degree $d$ with
$|P|\le1$ on $\T$ such that $r(\lambda)=P(z_{\sigma,\gamma}(\lambda))$ on $\R$.
If $P$ is an ordinary polynomial of degree $d$, the corresponding $r$ lies in $\Rat_{d,d}(\{\sigma-i\gamma\})$.
\end{theorem}
\begin{proof}
Write $t=\lambda-\sigma$ and $z=z_{\sigma,\gamma}(\lambda)=(t-i\gamma)/(t+i\gamma)$,
so that $$z^k=(t-i\gamma)^k(t+i\gamma)^{-k}= \frac{(t-i\gamma)^{d+k} (t+i\gamma)^{d-k}}{(t-i\gamma)^d (t+i\gamma)^d}$$ for every $k$, negative $k$ included.
Then  $P(z)=\sum_{k=-d}^{d}c_kz^k$ becomes
\[
P(z(\lambda))=\frac{N(t)}{(t^{2}+\gamma^{2})^{d}},\qquad
N(t)=\sum_{k=-d}^{d}c_k\,(t-i\gamma)^{d+k}(t+i\gamma)^{d-k}.
\]
Note that  $N(t)$ is a polynomial of degree at most
$2d$, while $t^{2}+\gamma^{2}=(\lambda-\sigma-i\gamma)(\lambda-\sigma+i\gamma)$
has roots equal to  $\sigma\pm i\gamma$. Hence $r=P( z_{\sigma,\gamma})\in \Rat_d(\sigma, \gamma)$, and the order of each pole 
drops below $d$ where $N$ vanishes.

The map $P\mapsto r=P( z_{\sigma,\gamma}(\lambda))$ is linear, and injective since
$z_{\sigma,\gamma}$, a M\"obius transformation, is invertible. Note that the space of Laurent polynomials of degree $d$ and that of rational functions with denominator  $((\lambda-\sigma)^2+\gamma^2)^d$ fixed and numerator of degree at most $2d$  have the same dimension, $2d+1$. An injective linear map between spaces of equal finite
dimension is a bijection, so every such $r$ arises from a unique $P$.
Since $z_{\sigma,\gamma}$ maps $\R$ onto $\T\setminus\{1\}$,  $\sup_\R|r|=\sup_\T|P|$, hence the bound
$|P|\le1$ and the bound $|r|\le1$ are the same constraint.

If only nonnegative powers appear the denominator is $(t+i\gamma)^d$,
and each numerator $(t-i\gamma)^k(t+i\gamma)^{d-k}$ has degree exactly $d$, giving
$r\in\Rat_{d,d}(\{\sigma-i\gamma\})$.
\end{proof}

\begin{remark}\label{rem:free-numerator}
The numerator $N$ in Theorem~\ref{thm:class} is free of degree up to $2d$. When its leading coefficients in the monomial basis vanish, $N$ has degree
strictly less than $2d$ and the rational is of type $(j,2d)$ with
$j<2d$. The shifted resolvent is the extreme case: at $d=1$ the choice $c_0=-c_1$ makes the
degree-one term of $N$ cancel, leaving
\[
c_0+c_1\,z_{\sigma,\gamma}(\lambda)=\frac{1}{\lambda-(\sigma-i\gamma)}
\quad\text{up to a constant,}
\]
a rational of type $(0,1)$. More generally every type $(j,d)$ over the denominator
$(\lambda-\sigma+i\gamma)^{d}$, and every type $(j,2d)$ over
$((\lambda-\sigma)^2+\gamma^2)^{d}$ with $j\le 2d$, is reached up to a positive scaling factor;
the numerator is chosen at will, subject only to the normalization $|r|\le1$ on $\R$ that GQSP
imposes, which fixes the overall scale but not the shape of $r$.

We remark that the freedom is in the {numerator}, not in the poles. The two poles are pinned at
$\sigma\pm i\gamma$; only their common order and the numerator are at the user's disposal.
A single conjugate pair therefore already gives a rich family: every rational function bounded
by one on $\R$ whose poles lie at $\sigma\pm i\gamma$, each of order at most $d$.
\end{remark}

The following Corollary shows the cost of using $\Cay_{\sigma, \gamma}(A)$ in association with GQSP to realize a rational function with prescribed poles.
\begin{corollary}\label{cor:cost}
Every $r\in\Rat_d(\sigma,\gamma)$ with $|r|\le1$ on $\R$ is applied with $2d$ controlled applications of $\Cay_{\sigma,\gamma}(A)^{\pm1}$ and one ancilla, with no parity restriction and no linear-combination overhead.
\end{corollary}

\subsection{Approximation: reduction to the circle}\label{sec:approx}
Step~3 of the pipeline requires a Laurent polynomial in the eigenphase
$z=e^{i\varphi}$, that is a trigonometric polynomial in $\varphi$, which is
what GQSP realizes. We therefore view $f$ on the unit circle rather than on the
spectrum, as $F:\T\to\C$,
\[
F(z) = f\bigl(\sigma-\gamma\cot(\varphi/2)\bigr), \qquad z = e^{i\varphi},
\]
the value of $f$ at the real number $\lambda(\varphi)$ carried to $z$. This
substitution reduces the problem into one of polynomial approximation on the circle, for
which Fourier truncation, Remez, Lawson iteration and linear programming are
available~\cite{EW76,Tang88,Watson80}.

\begin{proposition}\label{prop:reduction}
Let $S\supseteq\Lam(A)$ be an infinite compact set, let $f\in C^0(S)$, and let
$F(e^{i\varphi})=f(\sigma-\gamma\cot(\varphi/2))$. Then
\[
\min_{r\in\Rat_d(\sigma,\gamma)}\max_{\lambda\in S}|f(\lambda)-r(\lambda)| =\min_{P}\max_{z\in z_{\sigma,\gamma}(S)}|F(z)-P(z)| , \] where the minimum on the right is over the Laurent polynomials $P(z)=\sum_{k=-d}^{d}c_kz^k$.
\end{proposition}
If $F$ can be extended analytically to the annulus $\rho^{-1}<|z|<\rho$ with
$\rho>1$, a classical Bernstein-type theorem~\cite[Ch.~8]{Tr19} gives the
geometric decay of the error, $O(\rho^{-d})$, faster as the annulus is wider. Now
$F$ is $f$ seen through the M\"obius change of variable \eqref{eq:phasemap}, so
an annulus around the circle corresponds, under that map, to a region around
the real axis in the $\lambda$-plane, inside which $f$ must be analytic. The widest annulus one may
take is therefore limited by the singularity of $f$ closest to the axis, and
$\gamma$ has to be chosen accordingly.

Classically, rational approximation outperforms polynomial approximation on
functions such as the sign function or $1/\lambda$ near the origin because the
poles may be placed freely, and the optimal constructions, Zolotarev's among
them, cluster them geometrically towards the singularity. A single Cayley
factor pins its poles to the one conjugate pair $\sigma\pm i\gamma$, so by
Proposition~\ref{prop:reduction} the approximation is polynomial in the
transformed variable. Placing that pair next to the difficulty already helps, but Zolotarev's rate,
geometric with a ratio that depends only logarithmically on the gap, requires
several poles spaced geometrically, not just one.

One condition on $f$ is specific to this encoding. The angle map sends both
ends of the real axis to $z=1$, so $F$ is continuous on the circle only if the
two limits of $f$ at $\pm\infty$ agree. For a matrix of finite size this is no
restriction: the spectrum is bounded, its image is an arc that stays away from
$z=1$, and the values of $f$ outside the spectrum play no role, so any $f$
continuous on the spectrum can be approximated. The condition matters only for
spectra that grow without bound, as when a differential operator is discretized
on finer and finer grids: the arc then approaches $z=1$ and a function whose
two ends disagree, as for example the sign function,
cannot be followed uniformly.

\subsection{The shifted resolvent}\label{sec:resolvent}

For $w=\sigma-i\gamma$, a point  off the real axis, the resolvent
$g(\lambda)=\frac{1}{\lambda-\sigma+i\gamma}$
is rational of type $(0,1)$ with its only pole at $w$, and applying $g(A)$ to
a vector $b$ returns the solution of the shifted system
$(A-\sigma I+i\gamma I)x=b$, the basic step of shift-and-invert and rational
Krylov methods.

Hence, the shifted resolvent lies in the class of Theorem~\ref{thm:class} at degree one, and the
identity is exact:
\[
\frac{1-z_{\sigma,\gamma}(\lambda)}{2i\gamma}
=\frac{1}{\lambda-\sigma+i\gamma}=g(\lambda).
\]
The Laurent polynomial $p(z)=(1-z)/(2i\gamma)$ involves only $z^0$ and $z^1$,
so no approximation is made and $U^*$ is not needed.

GQSP requires the polynomial it implements to be bounded by one on $\T$, and
$\norm{p}_{\infty,\T}=1/\gamma$, attained at $z=-1$, the image of
$\lambda=\sigma$. The circuit therefore implements $P=\gamma p$ and the $\ket0$ branch of the ancilla carries
$\gamma\,g(A)b$, so recovering $g(A)b$ takes $O(1/(\gamma\norm{g(A)b}))$
rounds of amplitude amplification.

Every unitary-based method pays a factor of this kind. The circuit reads a
block of a unitary, which is a contraction, so a function of supremum
$1/\gamma$ can only appear subnormalized by $\gamma$; the polynomial approach
pays the same through the subnormalization of its own block encoding. Since $g$ is reached exactly at $d=1$, the gain of the  Cayley encoding is in
the degree, whereas a polynomial method must
first approximate it, at a degree growing like $1/\gamma$ because the pole
lies at distance $\gamma$ from the spectrum. 

\section{The complete family of rational encodings: products of Cayley factors}\label{sec:multipole}

The construction so far uses a single conjugate pair $\sigma\pm i\gamma$.
Multiplying Cayley factors places the poles at any finite set of points off
the real axis, still with one ancilla and no linear combination. The point of
this section is not the extension itself, which on the targets considered here
brings no advantage, but the proof that the resulting family is the complete
list of rational unitary encodings. 
\begin{definition}\label{def:multipole}
For $w_1,\dots,w_m\in\C\setminus\R$, with $q(\lambda)=\prod_j(\lambda-w_j)$ and $q^*(\lambda)=\prod_j(\lambda-\bar w_j)$, the \emph{multi-pole encoding} is
\[
\Cay_q(A)=q^*(A)\,q(A)^{-1}=\prod_{j=1}^m\bigl(A-\bar w_jI\bigr)\bigl(A-w_jI\bigr)^{-1}.
\]
\end{definition}

The factors of a Cayley encoding commute and are unitary, so
$\Cay_q(A)$ is a single unitary on the system register, with eigenvalues
$z_q(\lambda_k)$ where $z_q=q^*/q$.
%
The next theorem shows that every rational function unitary on every Hermitian matrix has this form, up to a unimodular constant.
\begin{theorem}\label{thm:completeness}
Let $r$ be rational, not identically zero, with $r(A)$ defined and unitary for every Hermitian $A$. Then $r=c\,q^*/q$ for a polynomial $q$ with no real roots and $|c|=1$; that is, $r(A)=c\,\Cay_q(A)$. Conversely every such $r$ has this property.
\end{theorem}
\begin{proof}
We prove the forward direction; the converse, that every such product is 
unitary on every Hermitian $A$, is immediate since each Cayley factor is.

Suppose $r$ is unitary on $A$ for every Hermitian $A$. Evaluating on 
eigenvalues we have $|r(\lambda)|=1$ for 
every real $\lambda$, and $r$ has no pole on $\R$ (a real pole would make 
$r(A)$ unbounded for some $A$). Write $r=p/q$ with $p,q$ coprime polynomials, 
and let $p^*(\lambda)=\overline{p(\bar\lambda)}$ denote the polynomial over $\C$ obtained 
by conjugating the coefficients. 

The condition $|r|=1$ on $\R$ means $|p(\lambda)|^2=|q(\lambda)|^2$ there, that 
is $p\,p^*=q\,q^*$ for all real $\lambda$; since two polynomials agreeing on all 
values in $\R$ are equal, this identity holds for all $\lambda$. Now $p$ divides the 
left-hand side $p\,p^*$, hence also the right-hand side $q\,q^*$; being coprime 
to $q$, it must divide $q^*$, i.e. $p\mid q^*$. The same argument with the roles of $p$ and $q$ 
exchanged gives $q^*\mid p$. Two polynomials that divide each other are equal up 
to a constant, so $p=c\,q^*$ with $|c|=1$ by the normalization $|r|=1$. 

Therefore $r=p/q=c\,q^*/q$, a unimodular constant times the ratio of a 
polynomial to its conjugate-reflection.
\end{proof}
\begin{proposition}\label{prop:multipole}
Let $w_j=\sigma_j-i\gamma_j$ with $\gamma_j>0$ for $j=1,\dots,m$, let
$q(\lambda)=\prod_j(\lambda-w_j)$ and $z_q=q^\ast/q$. The angle map of $z_q$
is the sum of the angle maps of its factors,
$\varphi_q=\sum_j\varphi_{\sigma_j,\gamma_j}$, so that
\begin{enumerate}
\item the accuracy density is
\[
\varphi_q'(\lambda)=\sum_{j=1}^{m}\frac{2\gamma_j}{(\lambda-\sigma_j)^2+\gamma_j^2},
\qquad \int_\R\varphi_q'=2\pi m ;
\]
\item degree-$d$ GQSP on $\Cay_q(A)$ reaches the functions
$\lambda\mapsto P(z_q(\lambda))$ with $P$ a Laurent polynomial of degree $d$
satisfying $|P|\le1$ on $\T$, using $2d$ applications of the encoding and
$2d+1$ angles, both independent of $m$. These functions lie in the space of
rational functions with denominator $(qq^\ast)^{d}$ and numerator of degree at
most $2md$, which has dimension $2md+1$, and span a subspace of it of dimension
$2d+1$.
\end{enumerate}
\end{proposition}
\begin{proof}
Eigenvalues of a product of commuting unitaries multiply, so the phases add:
$\varphi_q=\sum_j\varphi_{\sigma_j,\gamma_j}$. Differentiating termwise and
using $\int_\R\varphi_{\sigma,\gamma}'=2\pi$ (Section~\ref{sec:params}) for
each factor gives item~1.

For item~2, write $z_q^k=(q^\ast)^kq^{-k}$ and multiply numerator and
denominator by $q^{d}(q^\ast)^{d}$: the term of index $k$ becomes
$(q^\ast)^{k+d}q^{d-k}$, of degree $m(k+d)+m(d-k)=2md$, over the common
denominator $(qq^\ast)^{d}$, also of degree $2md$. Hence $P(z_q(\lambda))$ has
denominator $(qq^\ast)^{d}$ and numerator of degree at most $2md$. Distinct $P$
give distinct functions of $\lambda$, since $z_q$ is a nonconstant rational
function and therefore takes infinitely many values: if $P(z_q(\lambda))=0$ for
all $\lambda$, then $P$ has infinitely many roots and vanishes. The reachable
functions therefore span a subspace of dimension $2d+1$ of a space of dimension
$2md+1$.
\end{proof}
The counts of applications and angles are those of GQSP applied to the single
unitary $\Cay_q(A)$, and do not involve $m$; the cost of realizing one
application does, since $\Cay_q(A)$ is a product of $m$ Cayley factors.

The inclusion in item~2 is strict as soon as $m\ge2$, and this is what
separates the multi-pole family from the single pair. By
Remark~\ref{rem:free-numerator}, one conjugate pair already gives every bounded
rational function with those poles, each of order at most $d$, the numerator being free; with several
pairs the poles may be placed at will, but the numerator is no longer free,
being determined by the denominator through $z_q$.
The full type-$(2md,2md)$
class is reached instead by a linear combination of resolvents~\cite{SVCKV24},
at the cost of $\lceil\log_2 m\rceil$ ancillas and an $\ell_1$
subnormalization factor. By Theorem~\ref{thm:completeness} no rational unitary encoding, processed
alone by GQSP, reaches a free numerator: the wider class requires either
additional ancillas, as in the linear combination of resolvents, or a
different way of combining the encodings.

These products appear classically as well. For a dense $A$ the standard way of
applying $e^{-iA\tau}$ is not a polynomial method but an implicit one: the
Crank--Nicolson propagator is $-\Cay_{0,\gamma}(A)^{-1}$ with $\gamma=2/\tau$.
Higher-order variants are given by the diagonal Pad\'e approximants of $e^{z}$,
which have modulus one on the real line and are therefore again products of
Cayley factors, by Theorem~\ref{thm:completeness}. The same construction has
been proposed on a quantum computer, realizing those approximants through a
linear system solver~\cite{DLX22}; the present framework contains it as the
particular case in which the poles are fixed by the order of the approximant
rather than chosen in response to $f$.


\section{Comparison with the standard access models}\label{sec:threemodels}

In this section we compare our Cayley transform with the two standard
access models. Table~\ref{tab:comparison} present an overview of the main differences of the three access models. The block-encoding model embeds $A/\alpha$, with $\alpha\ge\norm A$, in the
leading block of a unitary $U_A$ acting on the system register and $a$ ancillas,
and transforms the singular values of the encoded block by
QSVT~\cite{GSLW19,MRTC21}. The matrix-exponential model supplies $e^{-iAt}$, a
unitary on the system register with no ancillas, processed by GQSP with one
ancilla to give arbitrary trigonometric polynomials in $\lambda t$. Each access model admits a natural transformation mechanism: GQSP for unitaries (acting on eigenvalues), QSVT for block encodings (acting on singular values). QSVT cannot be applied to unitaries because their singular values are all one.

The exponential and Cayley models fit the same template: a unitary $U$ commuting
with $A$, with $Uv=e^{i\varphi(\lambda)}v$, processed by degree-$d$ signal
processing to give $P(e^{i\varphi})$. They differ only in the substitution
$\lambda\mapsto\varphi$, so that choosing between them is choosing a change of
variable, and with it a classical approximation problem and an accuracy
density. The block-encoding model does not fit the template, since $U_A$ acts
on the ancillas as well and does not commute with $A$, but it enters the same
comparison through the Chebyshev change of variable $\lambda=\alpha\cos\varphi$,
under which the Chebyshev polynomials of the first kind satisfy
$T_k(\lambda/\alpha)=\cos k\varphi$.

\begin{table}
\small
\begin{tabular}{|@{}l|p{3cm}|p{3cm}|p{4.0cm}@{}|}
\toprule
Properties & block-encoding & mat. exponential & Cayley\\
\midrule
handle & $U_A$, $A/\alpha$ in a block & $e^{-iAt}$ & $\Cay_{\sigma,\gamma}(A)$\\
ancillas for handle & $a$ & none & none\\
substitution & $\varphi=\arccos(\lambda/\alpha)$ & $\varphi=-\lambda t$ & $\varphi=-2\arccot\frac{\lambda-\sigma}{\gamma}$\\
transformation & QSVT & GQSP & GQSP\\
reachable class & polynomials & trig.\ polynomials & $\Rat_d(\sigma, \gamma)$, poles $\sigma\pm i\gamma$ \\
classical analogue & Chebyshev & Fourier & prescribed-pole rational\\
accuracy density $\varphi'$ & $1/\sqrt{\alpha^2-\lambda^2}$ &
$t$, $t\le\pi/\alpha$ & $2\gamma/((\lambda-\sigma)^2+\gamma^2)$\\
deg. to distinguish $\Delta$ & $\alpha/\Delta$ away from $\pm\alpha$ &
$\alpha/\Delta$ & $\gamma/(2\Delta)$ near $\sigma$\\
needs $\norm A$ bound & yes & yes (aliasing) & no\\
target condition & spectrum in $[-\alpha,\alpha]$ & periodicity & cont. on the spectrum\\
\bottomrule
\end{tabular}
\caption{Comparison of the three access models.} \label{tab:comparison}
\end{table}

The Cayley column extends to $m$ pole pairs through $\Cay_q(A)$: the
substitution becomes a sum of $m$ arc-cotangents, the density a sum of $m$
bells, and the reachable class still has dimension $2d+1$
(Proposition~\ref{prop:multipole}).

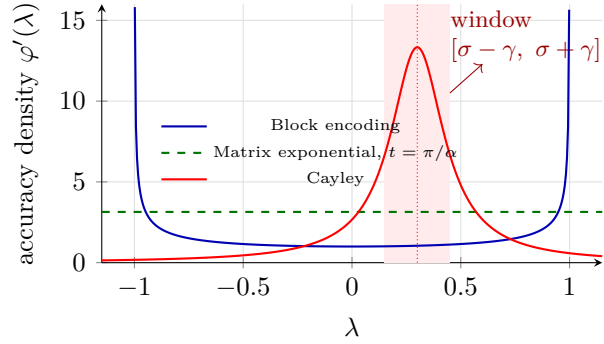
\begin{figure}[t!]
\centering
\begin{tikzpicture}
\begin{axis}[
    width=8.2cm, height=5cm,
    xlabel={$\lambda$},
    ylabel={accuracy density $\varphi'(\lambda)$},
    title={Accuracy density for the three access models ($\alpha=1$, $\sigma=0.3$, $\gamma=0.15$)},
    xmin=-1.15, xmax=1.15,
    ymin=0, ymax=16,
    legend pos=north west,
    legend style={
            font=\tiny,
            draw=none,
            fill=none,
            at={(0.1,0.60)},
            anchor=north west,
        },
    grid=major,
    grid style={gray!25},
    axis lines=left,
    tick label style={font=\small},
    label style={font=\small},
    title style={font=\small},
    clip=true,
]

\node[red!60!black, font=\footnotesize, align=left] at (axis cs:0.80,14)
    {window\\ $[\sigma-\gamma,\ \sigma+\gamma]$};
\addplot[fill=red!8, forget plot,  draw=none] coordinates {(0.15,0) (0.15,16) (0.45,16) (0.45,0)} \closedcycle;

\draw[->, red!60!black, thin] (axis cs:0.45,10.5) -- (axis cs:0.60,12.3);

\draw[red, densely dotted, thin] (axis cs:0.3,0) -- (axis cs:0.3,16);
 
\addplot[
    domain=-0.998:0.998, samples=400, samples y=1,
    thick, blue!70!black,
] {1/sqrt(1 - x^2)};
\addlegendentry{Block encoding}
 
\addplot[
    domain=-1.15:1.15, samples=2,
    thick, dashed, green!45!black,
] {3.14159};
\addlegendentry{Matrix exponential, $t=\pi/\alpha$} 
 
\addplot[
    domain=-1.15:1.15, samples=400,
    thick, red,
] {2*0.15 / ((x-0.3)^2 + 0.15^2)};
\addlegendentry{Cayley}

\end{axis}
\end{tikzpicture}

\caption{Accuracy density $\varphi'(\lambda)$ for the three access models, each
drawn at the parameters its model admits. The exponential density is constant
and is drawn at its ceiling $t=\pi/\alpha$, which the absence of aliasing
imposes; the block-encoding density is fixed by $A$ alone and rises towards the
spectral ends $\pm\alpha$; the Cayley density is peaked at $\sigma$ with value
$2/\gamma$ and half-width $\gamma$, and is the only one that can be placed.}
\label{fig:accuracy-density}
\end{figure}

The asymmetry between the three models is more profound than the row Table~\ref{tab:comparison} on
$\norm A$ suggests. In the first two models the unitary being processed, $U_A$ or $e^{-iAt}$, is
fixed by $A$ alone, and all the adaptation to $f$ happens in the choice of $P$. In fact, $\alpha$ is fixed by $A$, and $t$, although free, only scales the
whole spectrum's accuracy up or down together; it cannot shift accuracy toward
one part of the spectrum, the way $\sigma$ and $\gamma$ do. Its range is moreover
bounded, since $\varphi=-\lambda t$ sends the spectrum onto an arc of length at most
$2t\alpha$, with $\alpha\ge \norm A$ and distinct eigenvalues receive distinct phases only if that arc
fits in the circle, which forces $t\alpha\le\pi$. The exponential density is
therefore uniform and at most $\pi/\norm A$, where the Cayley density $2/\gamma$ is concentrated and unbounded as
$\gamma\to0$. The integral of $\varphi'$ over an interval is the length of the
arc onto which that interval is mapped, hence the number of eigenvalues a given
degree can separate there. Over the window $[\sigma-\gamma,\sigma+\gamma]$ the
Cayley encoding assigns half of the circle, against $2\gamma\pi/\alpha$ for the
exponential model at its ceiling, a ratio $\alpha/(2\gamma)$ that grows without
bound as the width narrows (Figure~\ref{fig:accuracy-density}).

\section{The cost of  realizing the Cayley encoding}\label{sec:cost}
In Sections~\ref{sec:cayley}, \ref{sec:multipole} and~\ref{sec:threemodels} we
worked in the query model: $\Cay_{\sigma,\gamma}(A)$ is treated as an oracle, and
the reachable class and the approximability of $f$ do not depend on how the
oracle is built. In that model the cost is the number of queries to the unitary that encodes $A$ and to its inverse, which is proportional to the degree. By
\eqref{eq:ddouble}, distinguishing eigenvalues at distance $\Delta$ near $\sigma$
requires a degree of order $\gamma/\Delta$ in the Cayley class, against
$\alpha/\Delta$ in the two standard models (Table~\ref{tab:comparison}). The
Cayley encoding therefore needs a lower degree when the difficulty of $f$ is
confined to a region of width comparable to $\gamma$, small compared with
$\alpha$, and for the shifted resolvent the degree is one, against a degree
growing as $1/\gamma$ for a polynomial method (Section~\ref{sec:resolvent}). When
the variation of $f$ is spread over the whole spectral set there is nothing to
concentrate, and the advantage disappears. The realization model accounts
instead for the cost of one call. We show that, starting from a block encoding of
a Hermitian matrix, we can construct a block encoding of the Cayley encoding at a
cost that we make precise below.

\paragraph{A generic realization from a block-encoding.}
Given an $\alpha$-block-encoding $U_A$ of $A$, with $\alpha\ge\norm A$, the
resolvent identity
\[
\Cay_{\sigma,\gamma}(A)=I-2i\gamma R,\qquad R=M^{-1},\quad M=A-\sigma I+i\gamma I,
\]
expresses the encoding through the shifted resolvent. The shifted matrix $M$ is
obtained from $U_A$ as a linear combination of $A$ and $I$, and since
$\norm M\le\norm A+|\sigma|+\gamma$ the resulting block encoding has
subnormalization
$\alpha_M=\alpha+|\sigma|+\gamma$.
$M$ is normal, with smallest singular value
$s_{\min}(M)=\min_j\sqrt{(\lambda_j-\sigma)^2+\gamma^2}\ge\gamma$ and, by
\eqref{eq:cond}, condition number at most
$\sqrt{\norm{A-\sigma I}^2+\gamma^2}/\gamma$, of order $\norm{A-\sigma I}/\gamma$
for $\gamma\ll\norm{A-\sigma I}$. 

\begin{lemma}\label{lem:onecall}
Let $U_A$ be an exact  $(\alpha, a)$-block-encoding of $A$ and let
$\alpha_M=\alpha+|\sigma|+\gamma$. For every $\delta\in(0,1)$ there is a
circuit that is a $(1,a+O(1),\delta)$-block-encoding of
$\Cay_{\sigma,\gamma}(A)$, at a cost of $\Ot(\alpha_M/\gamma)$ queries to $U_A$
and $U_A^*$.
\end{lemma}

\begin{figure}[t]
\centering
\begin{tikzpicture}[x=1cm,y=1cm,>=stealth,scale=0.8,every node/.style={scale=0.8}]
\node[left] at (-0.5,2) {$\ket{0}$};
\draw (-0.3,2) -- (13.4,2);
\node[left] at (-0.5,1) {$|0\rangle^{\otimes a}$};
\draw[double distance=1.5pt] (-0.3,1) -- (13.4,1);
\node[left] at (-0.5,0) {$|b\rangle$};
\draw[double distance=1.5pt] (-0.3,0) -- (13.4,0);
\node[left] at (-0.5,3) {$\ket{0}$};
\draw (-0.3,3) -- (13.4,3);

\foreach \xx/\lab in {0.3/{\tiny $\phi_0$}, 1.7/{\tiny $\phi_1$}, 3.65/{\tiny $\phi_{n-1}$}, 4.8/{\tiny $\phi_n$}}{
  \draw (\xx,2) -- (\xx,1);
  \node[draw,circle,fill=white,minimum size=3pt,inner sep=0pt] at (\xx,1) {};
  \node[draw,fill=green!12,rounded corners,minimum width=0.1cm,minimum height=0.6cm] at (\xx,2) {\lab};
}
\foreach \xx/\lab in {1.0/{\small$U_A$}, 2.4/{\small$U_A^{*}$}, 4.25/{\small$U_A^{*}$}}{
  \node[draw,fill=green!10,minimum width=0.2cm,minimum height=2.05cm] at (\xx,0.5) {\lab};}
\node[fill=white] at (2.6,2) {$\cdots$}; \node[fill=white] at (3.15,0) {$\cdots$};\node[fill=white] at (3.15,1) {$\cdots$};
\draw[decorate,decoration={brace,amplitude=5pt,mirror},thick] (0.15,-0.7) -- (5.15,-0.7)
  node[midway,below=6pt,align=center] {\scriptsize QSVT: resolvent\\[-1pt]
  \scriptsize $R=(A-\sigma I+i\gamma I)^{-1}$,\\[-1pt]
  \scriptsize degree $n_{\rm inv}=\Ot(\alpha_M/\gamma)$,\\[-1pt]
  \scriptsize subnorm.\ $\alpha_R=\Theta(1/\gamma)$};

\node[draw,fill=blue!7,rounded corners,minimum width=2.0cm,minimum height=2.50cm] at (7.0,0.9)
  {\scriptsize $\begin{array}{c}\text{affine}\\ I-2i\gamma R\end{array}$};
\draw[decorate,decoration={brace,amplitude=5pt,mirror},thick] (5.9,-0.7) -- (8.3,-0.7)
  node[midway,below=6pt,align=center] {\scriptsize subnorm.\ $\beta=O(1)$};

\node[draw,fill=purple!8,rounded corners,minimum width=1.8cm,minimum height=4.0cm] at (9.7,1.5)
  {\scriptsize $\begin{array}{c}\text{uniform}\\ \text{sing.-value}\\ \text{amplification}\end{array}$};
\draw[->,thick] (10.9,0.75) -- (11.5,0.75);
\node[fill=white,right,align=left] at (11.5,0.50) {\scriptsize block-enc.\\[-1pt]\scriptsize of $\Cay_{\sigma,\gamma}(A)$};
\draw[decorate,decoration={brace,amplitude=5pt,mirror},thick] (8.8,-0.7) -- (10.6,-0.7)
  node[midway,below=6pt,align=center] {\scriptsize amplification\\[-1pt]
  \scriptsize error $\delta$};
\draw[red!50, densely dashed, thick] (5.5,-0.9) -- (5.5,3.5);
\draw[red!50, densely dashed, thick] (8.35,-0.9) -- (8.35,3.5);
\end{tikzpicture}
\caption{Realizing the Cayley transform from a block encoding $U_A$ of $A$.
Stage~(i), QSVT applied to $U_A$ with the bounded approximation of $1/x$,
produces a block encoding of the resolvent $R$ with subnormalization
$\alpha_R=\Theta(1/\gamma)$. Its degree $n_{\rm inv}$ is proportional to the condition
number of the normalized shifted matrix, hence $n_{\rm inv}=\Ot(\alpha_M/\gamma)$, and
this is the query cost of one application of the encoding. Stage~(ii) forms $I-2i\gamma R$, whose
subnormalization is $\beta=O(1)$. Stage~(iii) amplifies the singular values,
all equal to $1/\beta$, back to one, returning a genuine unitary at the cost of
one flag qubit and a polynomial of degree $O(\log(1/\delta))$.}
\label{fig:cayley-from-be}
\end{figure}
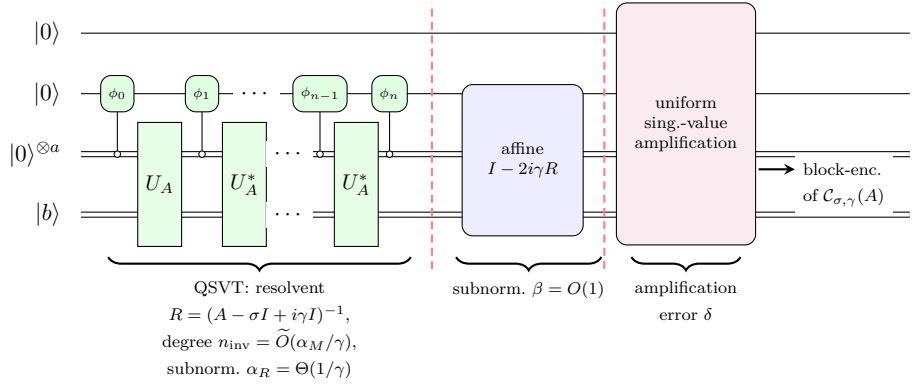

\begin{proof}
The construction proceeds in three stages, shown in
Figure~\ref{fig:cayley-from-be}.
\begin{description}
\item{(i)} The inversion of $M$, obtained by the  polynomial approximation of
$1/x$ of~\cite{CKS17}, and  implemented through QSVT~\cite{GSLW19,LT20a}, produces a block encoding of $R$ at a cost
of $\Ot(\alpha_M/\gamma)$ queries to $U_A$. Indeed the degree of the inversion
polynomial is proportional to the condition number of the normalized matrix
$B=M/\alpha_M$, whose smallest singular value $s_{\min}(M/\alpha_M)$ is at least $\gamma/\alpha_M$.
The block encoding of $R$ just constructed has, besides its query cost, a
subnormalization of its own, and it is this second quantity that matters next:
it alone governs how much amplification the following stage requires.  The inversion polynomial realizes
$p_{\mathrm{inv}}(x)\approx c/x$ subject to $|p_{\mathrm{inv}}|\le1$, which
forces $c\le s_{\min}(B)$; taking the largest allowed value,
$c=\Theta(\gamma/\alpha_M)$, gives $p_{\mathrm{inv}}(B)=\Theta(\gamma)M^{-1}$,
so the block encodes $R$ with subnormalization $\alpha_R=\Theta(1/\gamma)$: the
factor $\alpha_M$, and with it the dependence on $\alpha$, cancels in the
product $c\,B^{-1}$. This $1/\gamma$ scaling is
the worst case, attained when an eigenvalue of $A$ lies at $\sigma$.

\item(ii)  $\Cay_{\sigma,\gamma}(A)=I-2i\gamma R$ is obtained as a linear
combination of $I$ and $R$ and gives subnormalization
$\beta=1+2\gamma\,\alpha_R=O(1)$. Indeed,   $2\gamma\Theta(1/\gamma)=O(1).$

\item(iii) The block encoding therefore contains
$\Cay_{\sigma,\gamma}(A)/\beta$, whose singular values are all equal to $1/\beta$, $\Cay_{\sigma,\gamma}(A)$ being unitary. A subnormalized block cannot
serve directly as the GQSP signal, so one further step is needed: applying
Theorem~30 of~\cite{GSLW19} with amplification factor $\beta$ maps $1/\beta$ to
$1$ with a polynomial of degree $O(\beta\log(1/\delta))=O(\log(1/\delta))$,
using one additional flag qubit, and returns a genuine unitary block equal to
$\Cay_{\sigma,\gamma}(A)$ up to error $\delta$. Since $\beta=O(1)$, this
amplification costs only a constant factor and a logarithmic term, giving the
stated $(1,a+O(1),\delta)$-block-encoding at a total cost of
$\Ot(\alpha_M/\gamma)$ queries.
\end{description}
\end{proof}

The inversion of $M$ carries the whole cost: the affine step adds two queries,
and the amplification a factor $O(\log(1/\delta))$ since $\beta=O(1)$.

We are ready to prove the final result.

\begin{theorem}[Block-encoding of $f(A)$]\label{thm:endtoend}
Let $A$ be Hermitian with $\Lam(A)\subseteq S$, let $U_A$ be an
$(\alpha, a)$-block-encoding of $A$, and let $\alpha_M$ be as in
Lemma~\ref{lem:onecall}. Let $\varepsilon\in(0,1)$, let $\eta>0$ and let $P$ be a Laurent polynomial of degree $d$ with $\norm{P}_{\infty,\T}\le1$ such that
\begin{equation}\label{eq:etadef}
\max_{\lambda\in S}\bigl|f(\lambda)-\eta\,P(z_{\sigma,\gamma}(\lambda))\bigr|
\ \le\ \frac{\varepsilon}{2}.
\end{equation}
Then the circuit obtained by running degree-$d$ GQSP on the encoding of
Lemma~\ref{lem:onecall} is an $(\eta,a+O(1),\varepsilon)$-block-encoding of
$f(A)$, at a cost of $\Ot(d\,\alpha_M/\gamma)$ queries to $U_A$ and
$U_A^*$.
\end{theorem}
\begin{proof}
Set $\delta=(\varepsilon/(8d\eta))^2$ and let $W$ be the circuit supplied by
Lemma~\ref{lem:onecall} at accuracy $\delta$ at a cost of
$\Ot(\alpha_M/\gamma)$ queries to $U_A$.For a unit vector $\psi$ we have
$W(\ket0\otimes\psi)=\ket0\otimes\widetilde C\psi+g$, where
$\widetilde C=(\bra0\otimes I)W(\ket0\otimes I)$ is the block of $W$, with
$\norm{\widetilde C-\Cay_{\sigma,\gamma}(A)}\le\delta$, and $g$ satisfies
$(\bra0\otimes I)g=0$ and $\norm g^{2}=1-\norm{\widetilde C\psi}^{2}\le2\delta$. Hence
\[
\bigl\|W(\ket0\otimes\psi)-\ket0\otimes\Cay_{\sigma,\gamma}(A)\psi\bigr\|
\le\delta+\sqrt{2\delta}\le2\sqrt\delta ,
\]
and the same bound holds for $W^{*}$ and $\Cay_{\sigma,\gamma}(A)^{*}$, and for the
controlled versions of both.

Let $V$ be the degree-$d$ GQSP circuit built on $W$, and $V_0$ the same circuit
with $W$ replaced by a unitary $W_0$ satisfying
$W_0(\ket0\otimes\psi)=\ket0\otimes\Cay_{\sigma,\gamma}(A)\psi$ for every $\psi$.
Then $V_0$ keeps the ancillas of the lemma in $\ket0$ and its block is
$P(\Cay_{\sigma,\gamma}(A))$. Both circuits are products of the same $2d+1$
rotations and of $2d$ controlled applications of the encoding or of its inverse,
so a telescoping sum over these applications gives
\[
\bigl\|(V-V_0)(\ket0\otimes\psi)\bigr\|\le 2d\cdot2\sqrt\delta=\frac{\varepsilon}{2\eta}
\]
for every unit vector $\psi$. Since the compression onto $\ket0$ does not
increase the norm,
$\norm{\eta P(\Cay_{\sigma,\gamma}(A))-\eta\,(\bra0\otimes I)V(\ket0\otimes I)}
\le\varepsilon/2$.

Since $\Cay_{\sigma,\gamma}(A)$ commutes with $A$ and shares its eigenvectors
(Proposition~\ref{prop:cayley}), by~\eqref{eq:etadef}
\[
\bigl\|f(A)-\eta P(\Cay_{\sigma,\gamma}(A))\bigr\|
=\max_{\lambda\in\Lam(A)}
 \bigl|f(\lambda)-\eta P(z_{\sigma,\gamma}(\lambda))\bigr|
\le\frac{\varepsilon}{2},
\]
and the triangle inequality gives
$\norm{f(A)-\eta\,(\bra0\otimes I)V(\ket0\otimes I)}\le\varepsilon$.
The ancillas are those of Lemma~\ref{lem:onecall} together with the single GQSP
ancilla, and the query count is $2d$ times that of the lemma, with
$\log(1/\delta)=2\log(8d\eta/\varepsilon)$ absorbed in $\Ot$.
\end{proof}

Applying the block encoding of $f(A)$ to $\ket0\ket b$ gives the outcome $0$ on the ancilla with
probability $\norm{f(A)b}^2/\eta^2$, so the vector is
extracted by amplitude amplification, using a lower bound $\ell$ for
$\norm{f(A)b}$ to fix the number of rounds.

\begin{corollary}\label{cor:endtoend-vector}
Assume the hypotheses of Theorem~\ref{thm:endtoend}, let $b$ be a unit vector
and let $\ell>0$ satisfy $\ell\le\norm{f(A)b}$. Then for every
$\varepsilon\in(0,1)$ the normalized state $f(A)b/\norm{f(A)b}$ is prepared to
accuracy $\varepsilon$ at a cost of
\[
\Ot\!\left(\frac{d\,\alpha_M}{\gamma}\cdot\frac{\eta}{\ell}\right)
\]
queries to $U_A$.
\end{corollary}
\begin{proof}
Run the circuit $V$ of Theorem~\ref{thm:endtoend} at accuracy
$\varepsilon'=\varepsilon\ell/4$, which by the theorem costs
$\Ot(d\,\alpha_M/\gamma)$ queries, the dependence on $\varepsilon'$ being
logarithmic and absorbed in $\Ot$. Applied to $\ket0\ket b$, the circuit leaves
in the $\ket0$ branch a vector $w$ with
$\norm{w-\eta^{-1}f(A)b}\le\varepsilon'/\eta$, hence
$\norm w\ge \frac{\ell}{\eta}-\frac{\varepsilon'}{\eta}\ge \frac{\ell}{2\eta}$, using
$\varepsilon<1$. For two nonzero vectors,
\[
\Bigl\|\frac{w}{\norm w}-\frac{u}{\norm u}\Bigr\|
\ \le\ \frac{2\norm{w-u}}{\max\{\norm w,\norm u\}} ,
\]
so with $u=\eta^{-1}f(A)b$ the normalized vectors differ by at most
$2(\varepsilon'/\eta)/(\ell/(2\eta))=\varepsilon$. Fixed-point amplitude
amplification~\cite{YLC14}, calibrated on the lower bound $\ell/(2\eta)$ for
the amplitude, brings the $\ket0$ branch to weight $\Theta(1)$ in
$O(\eta/\ell)$ rounds, each one application of $V$ and of $V^*$, which
multiplies the count by that factor.
\end{proof}
A lower bound $\ell$ is available whenever $f$ does not vanish on the spectrum,
since $\norm{f(A)b}\ge\min_{\lambda\in\Lam(A)}|f(\lambda)|$ for a unit vector $b$.
For a spectral projector $\Pi$ the minimum is zero, and $\norm{\Pi b}$, the norm of
the component of $b$ in the projected subspace, can be arbitrarily small. It is
then estimated rather than used to extract $\Pi b$

\paragraph{Comparison with the matrix-exponential model.}
The matrix-exponential model admits the same construction from a block encoding
of $A$, at a cost of $\Ot(\alpha t)$ queries per application by
qubitization~\cite{LC17,LC19}. The two costs have the same form, $\Ot(\alpha t)$ against $\Ot(\alpha_M/\gamma)$,
whenever $|\sigma|\le\alpha$ and $\gamma\le\alpha$, so that $\alpha_M\le3\alpha$,
which is the case of interest. The evolution
time therefore plays the role of the reciprocal of the width: in both models
the parameter that concentrates the accuracy is the one that makes one
application expensive.
The two parameters differ in what they return: the exponential density is
uniform and bounded (Section~\ref{sec:threemodels}), where $2/\gamma$ is
unbounded and concentrated where it is placed.

\subsection{When does the Cayley encoding help?}\label{sec:budget}

In the query model the Cayley class reaches a prescribed accuracy at a degree
between three and four times smaller than the polynomial one, on the functions
whose difficulty is localized in part of the spectrum. We now weigh that gain
against the cost of one application, and find that the two cancel.

Let $\Delta$ be the separation fixed by $f$ as in Section~\ref{sec:params}. For
$1/\lambda$ on $[1/\kappa,1]$ $\Delta$ is $1/\kappa$, the modulus of the smallest eigenvalue: near that point the
function halves over an interval of that length, while near $\lambda=1$ it
varies far more slowly. Distinguishing two eigenvalues a distance $\Delta$
apart near $\sigma$ requires a degree $d\gtrsim\gamma/(2\Delta)$
by~\eqref{eq:resolution}, so the width that concentrates the accuracy is also
what raises the degree, while by Lemma~\ref{lem:onecall} a small width is what
makes one application expensive. 
\begin{corollary}\label{cor:cancellation}
Let $\Delta$ be the separation fixed by $f$ as in Section~\ref{sec:params}, and
let the degree be the smallest one meeting the resolution requirement, that is
$d=\Theta(\gamma/\Delta)$. Then the count
of Theorem~\ref{thm:endtoend} is $\Ot(\alpha_M/\Delta)$, independent of
$\gamma$, and that of Corollary~\ref{cor:endtoend-vector} is
$\Ot(\alpha_M\eta/(\Delta\,\ell))$, in which $\gamma$ appears only through
$\eta$.
\end{corollary}
The result of Corollary~\ref{cor:cancellation} is negative: what is saved in
degree is spent, at the same rate, in the shifted solve that each application
requires. When $|\sigma|\le\alpha$ and $\gamma\le\alpha$, so that
$\alpha_M\le3\alpha$, the resulting count $\Ot(\alpha_M/\Delta)$ is of the same
order as that of the polynomial method, which by Table~\ref{tab:comparison}
needs a degree $\alpha/\Delta$.
 This is where the quantum setting differs from the classical one. In the classical case the cost of a nearby pole is paid once, since the factorization of $A-wI$ is
computed and then reused at every application, so the reduction in degree is a net gain and shift-and-invert is amortized over the iterations. In the quantum setting, on the contrary,
each application of the encoding is a fresh solve costing
$\Ot(\alpha_M/\gamma)$, so the nearby pole reduces the degree and raises the
cost of one application at the same rate.

The question therefore becomes when the encoding can be realized more cheaply
than the construction above allows. Writing $C_{\mathrm{Cay}}$ for the cost of
one application in elementary gates, $C_{U_A}$ for the cost of one query to $U_A$,  the
block encoding of $A$, and $d$ and $n$ for the degrees at which the two classes
reach the same accuracy, the Cayley approach is the cheaper one when
$2d\,C_{\mathrm{Cay}}<n\,C_{U_A}$. Corollary~\ref{cor:cancellation} says that
the generic construction, for which
$C_{\mathrm{Cay}}=\Ot(\alpha_M/\gamma)\,C_{U_A}$, makes the two sides agree in
order, so the advantage must come from a realization that bypasses it
altogether. Tuning $\gamma$ does not help. For example, for $\gamma>\norm{A-\sigma I}$ the
resolvent has a geometrically convergent Neumann series and the solve is cheap,
but by~\eqref{eq:ddouble} such a width requires a degree of order
$\norm{A-\sigma I}/\Delta$, that of the polynomial methods, so the cheap and the
useful values of $\gamma$ do not overlap. The advantage must come from structure in $A$ that makes the shifted solve fast.

\paragraph{The case $f=1/\lambda$.}  Take $\norm A=1$, so that the spectrum lies in
$[-1,-1/\kappa]\cup[1/\kappa,1]$, and $\sigma=0$. The
pole of $1/\lambda$ lies on the real axis, so the function is not in the reachable
class; at degree one the encoding gives $1/(\lambda+i\gamma)$, with relative
error at most $\gamma\kappa$ on the spectrum. A small error therefore requires
$\gamma\lesssim1/\kappa$, and each application costs $\Ot(\kappa)\,C_{U_A}$.
Balancing degree and width in~\eqref{eq:ddouble} gives a total count of
$\Ot(\kappa)\,C_{U_A}$, the same order as the polynomial method, as
Corollary~\ref{cor:cancellation} predicts. There are two possible ways of realizing $\Cay_{\sigma,\gamma}(A)$ at a lower cost:
keeping the resolvent form $\Cay_{\sigma,\gamma}(A)=I-2i\gamma R$ and encoding $R$
directly from a structured description of it, so that $U_A$ is never formed,
which is the subject of Section~\ref{sec:structured}; or dropping that form and
realizing $\Cay_{\sigma,\gamma}(A)$ as a unitary circuit built from a
parametrization of the unitary itself, which is the approach of Section~\ref{sec:cmv}.

\subsection{Structured families}\label{sec:structured}

Encoding the shifted resolvent directly requires two properties of the matrix:
that it can be described by few parameters, and that those parameters can be
loaded as quantum states in polylogarithmic depth. If $A$ $A$ is real symmetric, tridiagonal and irreducible, the first property always holds, and the second holds in cases of
interest that we describe below. Note that $M=A-\sigma I+i\gamma I$ is then is complex symmetric and  tridiagonal as well, and by the
nullity theorem~\cite{VVM08} its inverse $R=M^{-1}$ is semiseparable of the
one-pair kind, that is, can be described by two complex generators $u$ and $v$ as
$R=\mathrm{triu}(uv^{T})+\mathrm{tril}(vu^{T},-1)$. Although $R$ is dense, a block encoding $U_R$ of $R$ can be built from its
generators alone, by an explicit circuit of depth $O(\log^3N)+D_{\mathrm{sp}}$,
where $D_{\mathrm{sp}}$ is the depth of the routines that prepare $u$ and $v$ as
quantum states~\cite{ADV26}. $U_A$
is never formed and the construction of Section~\ref{sec:cost} is bypassed; the
affine step $\Cay_{\sigma,\gamma}(A)=I-2i\gamma R$ and the amplification that
follows do not depend on how $U_R$ was obtained. The second property, that the
two generators can be prepared as quantum states in polylogarithmic depth, is not
automatic, and is what selects the subclass on which this approach is effective.

The encoding of~\cite{ADV26} gives $U_R$ with subnormalization
$\alpha_R=2\sqrt N\,\|u\|_2\|v\|_2$, so the affine step
$\Cay_{\sigma,\gamma}(A)=I-2i\gamma R$ has subnormalization
$\beta=1+2\gamma\alpha_R=1+4\gamma\sqrt N\,\|u\|_2\|v\|_2$, and $O(\beta)$ rounds of
amplitude amplification recover the unitary. One application of the encoding
therefore has depth $O(\beta(\log^{3}N+D_{\mathrm{sp}}))$, independent of
$\kappa(M)$, against $O(\log^{2}N)$ for one query to $U_A$ when $A$ is
tridiagonal~\cite{SCC24}. When the generators are prepared at polylogarithmic
depth, as discussed below, one application of the Cayley encoding is thus more
expensive than one query to $U_A$.
The cost of computing $f(A)b$, however, is the cost of one application multiplied
by the number of applications, which is fixed by the degree of the approximant
and by the rescaling $\eta$. Writing $n$ for the degree at which the polynomial
class reaches the target accuracy and $d$ for the degree in the Cayley class, the
more expensive applications are compensated by their smaller number, and the Cayley approach is the cheaper one, up to the logarithmic factor, when
$2d\,\beta\,\eta\lesssim n\,\eta_{\mathrm{pol}}$, with $\eta_{\mathrm{pol}}$ the
rescaling of the polynomial approximant, which equals $2$ for the inversion by
QSVT. Decreasing $\gamma$ lowers $\beta$ but raises $\eta$,
so the product $\beta\eta$ attains its minimum at an intermediate value of
$\gamma$, depending on the required accuracy.

The critical ingredient is the state preparation of the generators. Loading an
arbitrary vector of $N$ entries takes $\Theta(N)$ gates, so $D_{\mathrm{sp}}$ is
polylogarithmic only when $u$ and $v$ admit a low-complexity description. For
the three-point discretization of the Sturm--Liouville operator $-(p(x)w')'$
on $(0,1)$ with Dirichlet conditions, the generators of the inverse are
discretizations of smooth functions, analytic when $p$ is, and can be prepared
at polylogarithmic depth by a polynomial transformation of the grid
variable~\cite[Section~5.1]{ADV26}. With the shift $\gamma=O(\lambda_{\min})$
the generators of $R=(A+i\gamma I)^{-1}$ are discretizations of the solutions of
the shifted problem, and the same construction applies.

On the Sturm--Liouville family the Cayley approach is advantageous when the
difficulty of $f$ lies at the scale of the smallest eigenvalue, as for
$f(\lambda)=\Delta/\lambda$ with $\Delta=\lambda_{\min}=\Theta(N^{-2})$. Taking
$\gamma=\Theta(\Delta)$, the degree is $d=O(\log(1/\varepsilon))$
by~\eqref{eq:ddouble}, and since $\norm R\le1/\gamma$ the subnormalization
satisfies $\beta\le1+4b_{u, v}\sqrt N$, with $b_{u, v}=\|u\|_2\|v\|_2/\norm R$. If $b_{u, v}$ and $\eta$
remain bounded, as we verify in Section~\ref{sec:numerics-jacobi}, the Cayley
approach costs $\Ot(\sqrt N)$, against $\Ot(N^{2})$ for QSVT applied to $U_A$ and
for the matrix-exponential model, whose uniform accuracy density requires
$\Ot(1/\Delta)$ applications.

\subsection{The reverse direction: structured unitaries}\label{sec:cmv}

One can also proceed in the opposite direction: start from a unitary $U$ that
is cheap to realize as a circuit and ask which Hermitian $A$ has it as its
Cayley transform. A natural candidate is the class of CMV matrices, the
pentadiagonal unitaries obtained as the product of two layers of $2\times2$
unitary blocks on consecutive, offset index pairs, parametrized by Schur
coefficients $\alpha_k\in\C$, $|\alpha_k|<1$~\cite{SimonOPUC}.

CMV matrices are a particular instance of the quasiseparable matrices of order
one~\cite{BDG20}, those for which
$\mathrm{rank}\,M(k\!:\!N,\,1\!:\!k-1)=\mathrm{rank}\,M(1\!:\!k-1,\,k\!:\!N)=1$
for every $k$,
that is, whose submatrices lying strictly below and strictly above the diagonal
have rank one, the diagonal being unconstrained. The class contains the
semiseparable-plus-diagonal matrices, obtained by adding a diagonal to a matrix
all of whose strictly triangular submatrices come from a single pair of
generators.

\begin{proposition}\label{prop:cmv-semisep}
Let $U=\Cay_{\sigma,\gamma}(A)$ be a CMV matrix with $1\notin\Lam(U)$. Then $A$ is Hermitian and one-pair semiseparable plus diagonal.
\end{proposition}
We use this construction starting from a semiseparable plus
diagonal $A$, and computing the coefficients $\alpha_k$  from it by the Szeg\H{o} recursion~\cite{SimonOPUC},  and the proposition shows that the matrices reachable in this way are contained in that class.
\begin{proof}
Write $U=LM$, with $L=\bigoplus_{k\ \mathrm{even}}\Theta(\alpha_k)$ and
$M=1\oplus\bigoplus_{k\ \mathrm{odd}}\Theta(\alpha_k)\oplus1$ the two layers of
$2\times2$ blocks
$\Theta(\alpha)=\bigl(\begin{smallmatrix}\bar\alpha&\rho\\\rho&-\alpha\end{smallmatrix}\bigr)$,
$\rho=\sqrt{1-|\alpha|^2}$, and let $X=(I-U)^{-1}$, so that by
Definition~\ref{def:cayley}
\begin{equation}\label{eq:invcay}
A=\sigma I+i\gamma(I+U)X=(\sigma-i\gamma)I+2i\gamma X.
\end{equation}
Since $L$ is unitary, $I-U=L(L^*-M)$, and $T:=L^*-M$ is tridiagonal: the blocks
of $L^*$ occupy the index pairs of $L$, those of $M$ the pairs offset by one,
and the union of the two families spans all entries satisfying $|i-j|\le1$. Its
off-diagonal entries are the numbers $\rho_k=\sqrt{1-|\alpha_k|^2}$, all
nonzero, so $T$ is irreducible and by the nullity theorem~\cite{VVM08} its
inverse is one-pair semiseparable: there are $x,y$ with
$(T^{-1})_{ik}=x_iy_k$ for $i\ge k$. Moreover $X=T^{-1}L^*$.

Let $i>j$ and let $\{p,p+1\}$ be the pair of $L$ containing $j$. Since $L^*$ is
block diagonal, $(L^*)_{kj}=0$ unless $k\in\{p,p+1\}$, and $i>j\ge p$ gives
$i\ge p+1$, so $i\ge k$ for both values of $k$ and the one-pair formula applies
to each term:
\[
X_{ij}=\sum_{k}(T^{-1})_{ik}(L^*)_{kj}
      =x_i\bigl(y_p(L^*)_{pj}+y_{p+1}(L^*)_{p+1,j}\bigr)=:x_i\,w_j .
\]
The entries of $X$ below the diagonal are  those of a rank-one matrix,
and by~\eqref{eq:invcay} the same holds for $A$, with $u=2i\gamma x$ and
$\bar v=w$, the identity contributing only to the diagonal.

Finally $A$ is Hermitian: $U$ unitary gives $X^*=-XU$, whence $X+X^*=X(I-U)=I$,
so the Hermitian part of $X$ is $I/2$ and $2i\gamma(X-I/2)=2i\gamma X-i\gamma I$
is Hermitian; adding $\sigma I$ gives~\eqref{eq:invcay}.
\end{proof}

Classically $U=LM$ is a product of two layers of non-overlapping $2\times2$
rotations, a consequence of the Schur parametrization of unitary Hessenberg
matrices~\cite{Gragg86}. Writing $M=Z^* L'Z$, where $Z$ is the cyclic
shift $\ket k\mapsto\ket{k+1\bmod N}$, realized as an integer incrementer in
depth $O(\log^2N)$~\cite{ABDP26}, each layer couples index pairs differing in
the last bit alone and is therefore a rotation on that qubit whose angle is
selected by the remaining $n-1$: a multiplexed rotation, decomposed into
single-qubit rotations and CNOTs by the standard
construction~\cite{Mottonen2005}. This gives $U$, as an $n$-qubit
unitary with no ancilla and no subnormalization, without the necessity of the amplification  phase.

The cost lies entirely in the angles, and the condition is the one met in
Section~\ref{sec:structured} for the generators. A generic sequence
$\{\alpha_k\}$ requires $\Theta(N)$ gates, since specifying $N$ independent
numbers takes at least that many, so a polylogarithmic circuit requires the
sequence to admit a low-complexity description. As for the generators, this
holds when the coefficients are samples $\alpha_k=h(k/N)$ of a smooth function
$h$, prepared by a polynomial transformation of the grid
variable~\cite[Section~5.1]{ADV26}, or when they are stored in a
QRAM~\cite{giovannetti2008quantum}.
We do not know examples of semiseparable-plus-diagonal matrices of interest in
applications whose Cayley transform is a CMV matrix with coefficients of this
kind, so this construction is presented as a possible direction rather than as a
realized one.

\section{Numerical experiments}\label{sec:numerics}

In this section we compare the approximations that the three access models can
produce. For a function $f$ and a spectral set $S\supseteq\Lam(A)$ we compute
the best approximation of $f$ on $S$ in the maximum norm within the class
reachable by each model. We measure only the approximating power of the three
classes, without performing any quantum simulation, and combine the degrees
obtained with the counts of Section~\ref{sec:cost} to give a cost. All
functions are normalized so that $\|f\|_{\infty,S}=1$, and all spectral sets are
contained in $[-1,1]$, so that $\alpha=1$.

\subsection{Method}\label{sec:numerics-method}

Each model approximates $f$ in the span of $\nu=2d+1$ basis functions,
\[
z_{\sigma,\gamma}(\lambda)^k,\qquad e^{-ikt\lambda},\qquad k=-d,\dots,d,
\qquad\qquad T_k(\lambda/\alpha),\qquad k=0,\dots,2d ,
\]
for the Cayley, the exponential and the polynomial model respectively. The
spectral set is discretized by $\max\{800,8\nu\}$ Chebyshev points on each of the
intervals whose union is $S$, so that the ratio of nodes to basis functions never
falls below eight. For the Cayley model, whose basis functions oscillate rapidly
near $\sigma$, these nodes are augmented with the preimages of points equispaced
in the phase variable; since a maximum over a grid never exceeds the maximum over
$S$, the additional nodes can only make the measured error of this model larger.
We denote by $\lambda_1,\dots,\lambda_m$ the nodes and collect in
$B\in\C^{m\times\nu}$ the values of the basis functions at them.

The approximant is determined by the coefficient vector $c$ such that $(Bc)_j$ equals
$P(z_{\sigma,\gamma}(\lambda_j))$, $P(e^{-it\lambda_j})$ and $p(\lambda_j/\alpha)$
for the three models, with $P(z)=\sum_{k=-d}^{d}c_kz^k$ and
$p(x)=\sum_{k=0}^{2d}c_kT_k(x)$. Hence,  minimizing over $c$ is minimizing over the
whole class reachable by the model. The best approximation on $S$ solves
\begin{equation}\label{eq:minimax}
\min_{c\in\C^{\nu}}\ \max_{1\le j\le m}\ \bigl|(Bc)_j-f(\lambda_j)\bigr| ,
\end{equation}
but the cost also depends on the rescaling
\[
\eta=\max\Bigl\{1,\ \max_{z\in\T}|P(z)|\Bigr\}\quad\text{or}\quad
\eta=\max\Bigl\{1,\ \max_{x\in[-1,1]}|p(x)|\Bigr\},
\]
for the two unitary models and for the polynomial one respectively,
which~\eqref{eq:minimax} does not control. Indeed, where the image of $S$ does not
cover that domain, nothing holds the approximant down, and it can grow. We
therefore solve the family
\begin{equation}\label{eq:family}
\min_{c\in\C^{\nu}}\ \Bigl\|
\begin{bmatrix} B \\ \mu \widehat B \end{bmatrix} c -
\begin{bmatrix} f \\ 0 \end{bmatrix}
\Bigr\|_{\infty},
\qquad \mu\ge0,
\end{equation}
where $\widehat B$ collects the values of the basis functions on a
discretization of the domain where the constraint acts, and $f$ the values of $f$ at the nodes. The choice $\mu=0$
gives~\eqref{eq:minimax}, while larger values of $\mu$ trade accuracy on $S$ for a
smaller $\eta$. Among the members we keep the one minimizing the product of the
error on $S$ by $\eta$; since $\mu=0$ belongs to the family, the selected
approximant costs no more than the solution of~\eqref{eq:minimax}, and the
criterion does not require the tolerance to be fixed in advance. 

Each member of~\eqref{eq:family} is a linear Chebyshev approximation problem with
complex coefficients, for which several methods are available, among them
exchange algorithms of Remez type~\cite{Tang88}, linear
programming~\cite{Watson80} and Lawson's iteration~\cite{EW76}. We use the
latter, because it applies unchanged to the three bases and to spectral sets made
of several intervals, and because at every step it provides both a lower and an
upper bound for the minimax error, which we use as stopping criterion. For
weights $w$ with nonnegative entries of unit sum, let $c_w$ minimize
$\sum_jw_j|r_j(c)|^2$, where $r(c)$ is the residual of the member, and let
$r=r(c_w)$; then
\[
\Bigl(\sum_{j}w_j|r_j|^{2}\Bigr)^{1/2}\ \le\
\min_{c}\max_{j}\bigl|r_j(c)\bigr|\ \le\ \max_{j}|r_j| ,
\]
the lower bound because the weights sum to one, the upper one because $c_w$ is
admissible. The iteration is stopped when the two bounds agree to within a
relative $10^{-4}$, so the computed approximant is within that factor of the best
one and the reported degrees do not depend on the accuracy of the iteration.
Each step is a weighted least squares problem, solved through a QR factorization
with column pivoting of the weighted and column-equilibrated basis matrix; the
normal equations are not formed, since the three bases are Vandermonde-like and
severely ill conditioned at high degree.

By Corollary~\ref{cor:endtoend-vector} the result is extracted in $O(\eta/\ell)$
rounds of amplitude amplification, where $\ell$ is a lower bound for
$\norm{f(A)b}$ and depends on $f$, $A$ and $b$ but not on the encoding. It is
therefore the same for the three models and cancels in the comparison, leaving
the effective cost $2d\,\eta$, since an approximant of degree $d$ is applied with
$2d$ queries, by GQSP for the Cayley and exponential models and by QSVT for the
polynomial one.

QSVT realizes real polynomials of definite parity. For the test functions with
no symmetry we use the mixed formulation, which realizes the even and odd parts
separately and combines them at the cost of one further ancilla and a factor
two in the subnormalization; for the others the basis is restricted to the
Chebyshev polynomials of the corresponding parity.
The details of the discretizations and of the selection of $\mu$ are documented
in the code, which reproduces all the tables and figures of this
section~\cite{zenodo}.
\subsection{The query model}\label{sec:numerics-query}
We tested eight functions, normalized so that $\max_S|f|=1$, on spectral sets
contained in $[-1,1]$. Four have a difficulty localized in a small region: the
resolvent $\gamma_0/(\lambda-\sigma_0+i\gamma_0)$, with $\sigma_0=0.3$ and
$\gamma_0=0.05$, on $[-1,1]$; the inverse $1/(\kappa\lambda)$, with $\kappa=20$, on
$[-1,-1/\kappa]\cup[1/\kappa,1]$; the sign function on $[-1,-0.05]\cup[0.05,1]$;
and the indicator of $[0.15,0.45]$, for which $f(A)$ is a spectral projector, on
$[-1,1]$ with a neighbourhood of half-width $0.02$ of each endpoint removed. Two
have a branch point at the origin: $\sqrt\lambda$ on $[0.02,1]$, which leaves it
outside the spectral set, and $\sqrt{|\lambda|}$ on $[-1,-0.02]\cup[0.02,1]$,
which surrounds it. The last two, $e^{-\lambda}$ and $e^{-8i\lambda}$ on $[-1,1]$,
vary over the whole spectral set and serve as controls. The centre $\sigma$ of
the Cayley encoding is placed where the difficulty lies: at the real part of the
pole for the resolvent, at the centre of the window, at the left end of the
spectral set for $\sqrt\lambda$ and $e^{-\lambda}$, and at the origin for the
others. Table~\ref{tab:query} reports, for each model, the smallest $2d\,\eta$
over the degrees at which the error of the selected approximant falls below
$10^{-6}$, with $\eta$ in brackets.

On the four functions with a localized difficulty the Cayley class is the
cheapest, and it reaches the resolvent exactly at $2d=2$, since it contains it by
construction. It is also the cheapest on the two square roots, whether the
spectral set lies on one side of the branch point or surrounds it: on
$\sqrt\lambda$ it costs less than half of the exponential model and a sixth of
the polynomial one, and on $\sqrt{|\lambda|}$ the polynomial class does not reach
the tolerance within the degrees tested. On $e^{-\lambda}$ the variation is
spread over the whole spectral set, so there is nothing to localize, and the
polynomial class is the cheapest: Chebyshev approximation of an entire function
converges faster than geometrically, whereas the two unitary models approximate
on the circle and converge only geometrically. On $e^{-8i\lambda}$ the
exponential model is the cheapest, since its class contains the target, as the
Cayley class contains the resolvent. The appropriate model therefore depends on
the class to which $f$ belongs and on where its difficulty lies relative to the
spectral set.

\begin{table}[h]
\centering\small
\begin{tabular}{|@{}l|c|c|c@{}|}
\hline
functions & $2d\eta$ Cayley & $2d\eta$ matrix exp. & $2d\eta$ block enc.\\
\hline
resolvent, pole at $0.3-0.05i$ & $2$   \ ($1.00$) & $192$ \ ($1.00$) &  $512$ \ ($1.00$) \\ 
$1/\lambda$, $\kappa=20$       & $91$ \ ($2.86$) & $343$ \ ($1.79$) & -- \\
$\sign\lambda$, gap $\pm0.05$  & $64$ \ ($1.00$) & $192$ \ ($1.00$) & $256$ \ ($1.00$) \\
window on $[0.15, 0.45]$, gap $\pm0.02$       & $192$ \ ($1.00$) & -- & -- \\
\hline
$\sqrt\lambda$ on $[0.02,1]$   & $61$ \ ($2.54$) & $97$( $1.01$) &         $380$ ( $1.48$)\\
$\sqrt{|\lambda|}$, gap $\pm 0.02$      & $175$ \ ($1.82$) & $258$ \ ($1.01$) & -- \\
\hline
$e^{-\beta\lambda}$, $\beta=1$          & $56$ \ ($2.32$) & $26$ \ ($1.62$) & $16$ \ ($1.00$) \\
$e^{-it\lambda}$, $t=8$   & $101$ \ ($1.05$) & $12$ \ ($1.00$) & $48$ \ ($1.00$) \\ \hline
\end{tabular}
\caption{Effective cost at tolerance $10^{-6}$, defined in the text, with the
rescaling $\eta$ in brackets. A dash means the tolerance is not reached within
the degrees tested. For the polynomial class the cost includes the factor two of
the mixed formulation, charged on the functions of no definite parity.}
\label{tab:query}
\end{table}

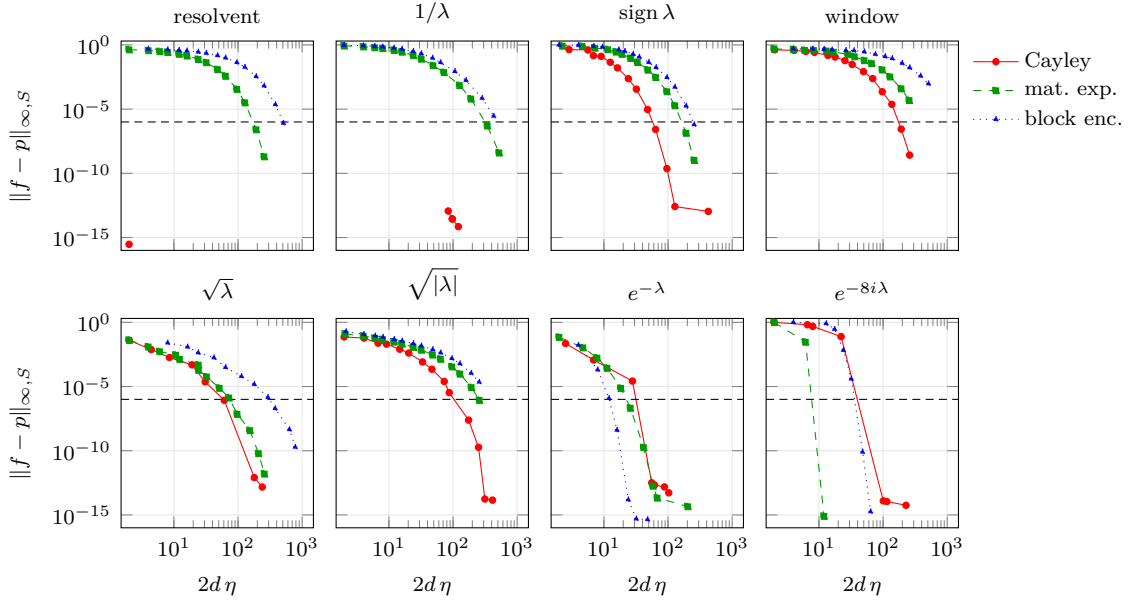
\begin{figure*}[t]
\centering
\begin{tikzpicture}
\begin{groupplot}[
  group style={group size=4 by 2, horizontal sep=0.30cm, vertical sep=0.9cm,
               ylabels at=edge left, yticklabels at=edge left,
               xlabels at=edge bottom, xticklabels at=edge bottom},
  width=0.275\textwidth, height=0.29\textwidth,
  xmode=log, ymode=log, xmin=1.5, xmax=1500, ymin=1e-16, ymax=2,
  xtick={10,100,1000}, ytick={1e-15,1e-10,1e-5,1},
  tick label style={font=\scriptsize}, label style={font=\scriptsize},
  title style={font=\scriptsize, yshift=-3pt},
  grid=major, grid style={line width=0.2pt, draw=gray!20},
  xlabel={$2d\,\eta$}, ylabel={$\|f-p\|_{\infty,S}$}]

\nextgroupplot[title={resolvent}]
\addplot[color=red, mark=*, mark size=1.2] coordinates {(2,2.99e-16) };
\addplot[color=green!60!black, mark=square*, mark size=1.2, dashed] coordinates {(2,4.29e-01) (4,3.66e-01) (6,3.12e-01) (8,2.67e-01) (12,1.95e-01) (16,1.43e-01) (24,7.61e-02) (32,4.26e-02) (48,1.24e-02) (64,3.78e-03) (96,3.58e-04) (128,3.19e-05) (192,2.55e-07) (256,2.01e-09) };
\addplot[color=blue, mark=triangle*, mark size=1.2, dotted] coordinates {(4,4.74e-01) (8,4.28e-01) (12,3.85e-01) (16,3.47e-01) (24,2.81e-01) (32,2.28e-01) (48,1.50e-01) (64,9.86e-02) (96,4.27e-02) (128,1.85e-02) (192,3.46e-03) (256,6.48e-04) (384,2.27e-05) (512,7.95e-07) };
\addplot[black, thin, densely dashed, forget plot] coordinates {(1.5,1e-6) (1500,1e-6)};

\nextgroupplot[title={$1/\lambda$}]
\addplot[color=red, mark=*, mark size=1.2] coordinates {(84.08,1.17e-13) (97.34,2.84e-14) (97.56,2.76e-14) (121,7.05e-15) };
\addplot[color=green!60!black, mark=square*, mark size=1.2, dashed] coordinates {(2,8.51e-01) (4,7.30e-01) (6,6.22e-01) (8,5.33e-01) (12,3.89e-01) (16,2.84e-01) (24,1.51e-01) (32,8.08e-02) (49.03,2.49e-02) (72.24,7.28e-03) (126.4,7.10e-04) (190.6,6.33e-05) (343.1,5.03e-07) (523.1,4.00e-09) };
\addplot[color=blue, mark=triangle*, mark size=1.2, dotted] coordinates {(2,9.50e-01) (4,8.60e-01) (6,7.78e-01) (8,7.04e-01) (12,5.76e-01) (16,4.72e-01) (24,3.16e-01) (32,2.12e-01) (48,9.51e-02) (64,4.27e-02) (107,8.61e-03) (159.6,1.74e-03) (284.4,7.05e-05) (431.5,2.87e-06) };
\addplot[black, thin, densely dashed, forget plot] coordinates {(1.5,1e-6) (1500,1e-6)};

\nextgroupplot[title={$\sign\lambda$}]
\addplot[color=red, mark=*, mark size=1.2] coordinates {(2.856,4.28e-01) (5.617,4.04e-01) (6.868,1.45e-01) (9.034,1.29e-01) (12.54,4.47e-02) (16.26,1.62e-02) (24.06,2.34e-03) (32.01,3.58e-04) (48,9.34e-06) (64,2.62e-07) (96,2.34e-10) (128,2.58e-13) (424.9,1.09e-13) };
\addplot[color=green!60!black, mark=square*, mark size=1.2, dashed] coordinates {(2.245,8.35e-01) (4.587,7.86e-01) (7.213,5.83e-01) (11.5,4.37e-01) (15.44,2.87e-01) (19.09,1.93e-01) (26.18,9.07e-02) (33.42,4.42e-02) (48.54,1.13e-02) (64.19,3.02e-03) (96.02,2.34e-04) (128,1.90e-05) (192,1.37e-07) (256,1.05e-09) };
\addplot[color=blue, mark=triangle*, mark size=1.2, dotted] coordinates {(2,9.82e-01) (4,9.49e-01) (6.928,7.67e-01) (9.477,6.77e-01) (17.12,4.27e-01) (21.16,3.23e-01) (28.55,1.90e-01) (35.66,1.14e-01) (50.1,4.38e-02) (65.12,1.74e-02) (96.28,2.94e-03) (128.1,5.21e-04) (192,1.76e-05) (256,6.23e-07) };
\addplot[black, thin, densely dashed, forget plot] coordinates {(1.5,1e-6) (1500,1e-6)};

\nextgroupplot[title={window},
  legend style={font=\scriptsize, at={(1.03,1)}, anchor=north west,
                draw=none, fill=none, cells={anchor=west}}]
\addplot[color=red, mark=*, mark size=1.2] coordinates {(2.031,4.14e-01) (4.108,3.73e-01) (6.27,3.13e-01) (8.504,2.58e-01) (13.82,1.52e-01) (17.82,1.14e-01) (25.43,5.97e-02) (32.97,3.03e-02) (49.55,8.25e-03) (68.92,2.37e-03) (97.1,2.24e-04) (137.1,2.36e-05) (192,2.77e-07) (259.8,2.63e-09) };
\addplot[color=green!60!black, mark=square*, mark size=1.2, dashed] coordinates {(2,4.85e-01) (4,4.44e-01) (6.004,3.87e-01) (8.29,3.81e-01) (13.07,3.49e-01) (16.9,2.83e-01) (28.35,1.81e-01) (35.92,1.23e-01) (50.95,6.14e-02) (66.33,3.64e-02) (97.11,1.16e-02) (128.4,3.31e-03) (192.1,3.92e-04) (256,4.85e-05) };
\addplot[color=blue, mark=triangle*, mark size=1.2, dotted] coordinates {(4,4.97e-01) (8,4.85e-01) (12,4.62e-01) (16,4.21e-01) (26.4,3.73e-01) (35.29,3.62e-01) (51.02,2.94e-01) (75.91,1.88e-01) (108.1,1.26e-01) (139.3,8.79e-02) (199.2,3.74e-02) (260.4,1.70e-02) (385.5,3.85e-03) (512.5,9.49e-04) };
\addplot[black, thin, densely dashed, forget plot] coordinates {(1.5,1e-6) (1500,1e-6)};
\legend{Cayley, mat.\ exp., block enc.}

\nextgroupplot[title={$\sqrt\lambda$}]
\addplot[color=red, mark=*, mark size=1.2] coordinates {(2,4.03e-02) (4.433,7.48e-03) (8.505,1.80e-03) (19.18,4.87e-04) (30.75,2.29e-05) (60.96,8.54e-07) (179.4,8.14e-13) (238.8,1.49e-13) };
\addplot[color=green!60!black, mark=square*, mark size=1.2, dashed] coordinates {(2,4.11e-02) (4,1.24e-02) (6,5.36e-03) (10.7,2.86e-03) (12.3,1.34e-03) (24.15,4.82e-04) (24.2,1.90e-04) (32.43,5.76e-05) (51.22,7.62e-06) (72.27,1.33e-06) (97.19,7.12e-08) (153.2,3.83e-09) (208.9,6.13e-11) (258.5,1.59e-12) };
\addplot[color=blue, mark=triangle*, mark size=1.2, dotted] coordinates {(8,2.52e-02) (16.34,1.22e-02) (24,4.31e-03) (42.13,1.76e-03) (64.01,3.12e-04) (113.3,6.19e-05) (179.7,1.50e-05) (292.7,1.45e-06) (380.2,1.94e-07) (634.9,4.49e-09) (782.4,1.92e-10) };
\addplot[black, thin, densely dashed, forget plot] coordinates {(1.5,1e-6) (1500,1e-6)};

\nextgroupplot[title={$\sqrt{|\lambda|}$}]
\addplot[color=red, mark=*, mark size=1.2] coordinates {(2,7.27e-02) (4.089,5.92e-02) (6.763,2.39e-02) (9.157,2.00e-02) (14.75,8.09e-03) (20.4,4.04e-03) (33.41,8.20e-04) (46.87,2.22e-04) (72.87,2.49e-05) (87.24,3.34e-06) (174.8,2.43e-08) (249.8,1.85e-10) (313.8,1.74e-14) (413.3,1.43e-14) };
\addplot[color=green!60!black, mark=square*, mark size=1.2, dashed] coordinates {(2.076,1.24e-01) (4,8.61e-02) (6.067,5.83e-02) (8,4.62e-02) (12,2.99e-02) (16,2.10e-02) (24,1.17e-02) (32,7.05e-03) (48,2.94e-03) (64,1.38e-03) (96.19,3.51e-04) (128.4,9.66e-05) (193,8.62e-06) (257.5,8.68e-07) };
\addplot[color=blue, mark=triangle*, mark size=1.2, dotted] coordinates {(2.142,1.91e-01) (4.053,1.16e-01) (6.111,8.79e-02) (8.095,6.93e-02) (12.1,4.80e-02) (16.09,3.58e-02) (24.08,2.23e-02) (32.06,1.50e-02) (48.32,7.66e-03) (64.24,4.28e-03) (96.12,1.53e-03) (128.1,6.05e-04) (192,1.10e-04) (256,2.23e-05) };
\addplot[black, thin, densely dashed, forget plot] coordinates {(1.5,1e-6) (1500,1e-6)};

\nextgroupplot[title={$e^{-\lambda}$}]
\addplot[color=red, mark=*, mark size=1.2] coordinates {(2.524,2.23e-02) (6.914,1.16e-03) (27.83,2.66e-05) (55.64,3.19e-13) (61.28,2.15e-13) (87.91,1.52e-13) (102.7,5.29e-14) };
\addplot[color=green!60!black, mark=square*, mark size=1.2, dashed] coordinates {(2,6.99e-02) (4.717,1.07e-02) (7.845,1.69e-03) (11.2,2.74e-04) (18.35,7.48e-06) (25.88,2.12e-07) (41.67,1.81e-10) (58.06,1.82e-13) (68.02,2.14e-14) (203.5,4.55e-15) };
\addplot[color=blue, mark=triangle*, mark size=1.2, dotted] coordinates {(4,1.66e-02) (8,2.01e-04) (12,1.18e-06) (16,4.07e-09) (24,1.53e-14) (32,5.00e-16) (48,4.44e-16) };
\addplot[black, thin, densely dashed, forget plot] coordinates {(1.5,1e-6) (1500,1e-6)};

\nextgroupplot[title={$e^{-8i\lambda}$}]
\addplot[color=red, mark=*, mark size=1.2] coordinates {(2,1.00e+00) (6.619,6.47e-01) (8,4.81e-01) (22.17,7.67e-02) (100.8,1.25e-14) (114.2,1.11e-14) (228,5.67e-15) };
\addplot[color=green!60!black, mark=square*, mark size=1.2, dashed] coordinates {(2,1.00e+00) (6.172,2.94e-02) (12,8.08e-16) };
\addplot[color=blue, mark=triangle*, mark size=1.2, dotted] coordinates {(4,1.00e+00) (12.98,7.99e-01) (17.57,2.89e-01) (24.16,6.60e-03) (32,3.89e-05) (48,7.98e-11) (64,1.80e-15) };
\addplot[black, thin, densely dashed, forget plot] coordinates {(1.5,1e-6) (1500,1e-6)};

\end{groupplot}
\end{tikzpicture}
\caption{Maximum error on the spectral set against the effective cost
$2d\,\eta$; for the
polynomial model the cost includes the factor two of the mixed formulation on
the functions of no definite parity. For each model the curve gives the smallest error attained by an approximant of
cost at most $2d\,\eta$. The dashed line marks the tolerance $10^{-6}$: the
first point of each curve below it is the entry of Table~\ref{tab:query}.}
\label{fig:curves}
\end{figure*}

Figure~\ref{fig:curves} shows, for each model, the smallest error attainable at a
given effective cost $2d\,\eta$, so that the table can be read off the figure at
the tolerance $10^{-6}$ and the figure shows how the comparison changes with the
accuracy required. On the functions with a localized difficulty the Cayley
curve lies to the left of the other two at every accuracy below $10^{-3}$; on
$e^{-\lambda}$ and $e^{-8i\lambda}$ the polynomial and the exponential models are
the cheapest, as in the table.

\subsection{A family in which the encoding costs polylogarithmically}
\label{sec:numerics-jacobi}

In this section we consider the computation of $f(A)b$ for a tridiagonal $A$.
Classically this computation takes   $\Omega (N)$ operations, so a quantum
approach is of interest only if it costs less than that, when the comparison is
made in gates. We take $A$ to be the three-point
discretization of the Sturm--Liouville operator $-(p(x)w')'$ on $(0,1)$ with
Dirichlet conditions, normalized so that $\norm A=1$, with four coefficients,
$p\equiv1$, $p(x)=e^{2x}$, $p(x)=1+9x^{2}$ and $p(x)=1+\sin(2\pi x)/2$, all
positive and bounded on $[0,1]$. Then $\Delta=\lambda_{\min}(A)=\Theta(N^{-2})$
and $\kappa=\Theta(N^{2})$. For $p\equiv1$ the matrix is the discrete Laplacian, and hnece is diagonalized by the discrete sine
transform and fast inversion~\cite{TAWL21} applies.

We consider functions whose variation is concentrated near the smallest
eigenvalue, in an interval $[\Delta,c\Delta]$ with $c$ a moderate constant, and
which vary slowly elsewhere. Examples are the spectral projector onto the
eigenvalues below a threshold of order $\Delta$, whose trace is the number of
those eigenvalues; its smoothed version given by the logistic function
$1/(1+e^{(\lambda-\mu)/T})$, with $\mu$ and $T$ of order $\Delta$; the exponential
$e^{-s\lambda}$ with $s$ of order $1/\Delta$; the resolvent at a complex shift of order $\Delta$; and
the inverse $\Delta/\lambda$.\footnote{For the inverse a cheaper route is
available: since $A^{-1}$ is itself one-pair semiseparable, it can be
block-encoded directly from its generators~\cite{ADV26}, with effective cost
$2\sqrt N\,\Delta\|u_0\|_2\|v_0\|_2=\Theta(\sqrt N)$, of the same order as the
Cayley approach and smaller by a constant factor, since the Cayley class reaches
the inverse through the resolvent. That route gives the inverse only.}
Functions that vary on all scales between $\Delta$ and one, such as
$\sqrt\lambda$ on $[\Delta,1]$, or uniformly over the spectrum, such as
$e^{-\beta i\lambda}$, are excluded.

For such functions the comparison is between degrees. The polynomial approach
applies QSVT to $U_A$, whose depth is $O(\log^2N)$ for a tridiagonal
$A$~\cite{SCC24}, and the exponential approach applies $e^{-iAt}$ at
polylogarithmic depth~\cite{AGSBZ24}; both have a uniform accuracy density, and
distinguishing eigenvalues at distance $\Delta$ requires a degree of order
$\alpha/\Delta=\kappa$. The Cayley approach, with $\sigma=0$ and the encoding
realized from the generators of $R=(A+i\gamma I)^{-1}$ as in
Section~\ref{sec:structured}, needs by~\eqref{eq:ddouble} a degree of order
$\gamma/\Delta$, that is $O(\log(1/\varepsilon))$ for $\gamma=\Theta(\Delta)$,
independent of $N$. The price is paid in the cost of one application, of depth
$O(\beta(\log^3N+D_{\mathrm{sp}}))$, and since $\gamma\norm R\le1$,
\[
\beta=1+4\gamma\sqrt N\,b_{u, v}\,\norm R\le1+4b\sqrt N,\qquad
b_{u, v}=\frac{\|u\|_2\|v\|_2}{\norm R},
\]
for every width. The Cayley approach is therefore advantageous if the balance $b_{u, v}$ of the
generators remains bounded and the generators can be prepared at
polylogarithmic depth.

 The Cayley approach is therefore advantageous if the balance
$b_{u, v}$ of the generators remains bounded and the generators can be prepared at
polylogarithmic depth.

The depth of the state preparation is governed by the degree of a polynomial
reproducing the generators: following~\cite[Section~5.1]{ADV26}, if
$u_i=g(t_i)$ on the grid $t_i=-1+2i/(N-1)$ and $P$ is a polynomial of degree
$d_{\mathrm{sp}}$ approximating $g$ on $[-1,1]$, the state proportional to $u$ is
prepared by a circuit whose depth is proportional to $d_{\mathrm{sp}}\log N$.
Table~\ref{tab:jacobi} reports $d_{\mathrm{sp}}$, the smallest degree, among
multiples of four, for which the least squares Chebyshev approximation of the
normalized generator has relative error below $10^{-12}$.

\begin{table}[t]
\centering\small
\begin{tabular}{|l|r|r|r|r|r|r|r|r|}
\hline
 & & \multicolumn{2}{c|}{$b_{u, v}$} & \multicolumn{2}{c|}{$\beta$} & \multicolumn{2}{c|}{$d_{\mathrm{sp}}$} & \\
$p(x)$ & $N$ & $\gamma_1$ & $\gamma_2$ & $\gamma_1$ & $\gamma_2$ & $\gamma_1$ & $\gamma_2$ & $\kappa$ \\
\hline
$1$ & $64$ & $3.21$ & $3.23$ & $1.00$ & $11.3$ & $4$ & $12$ & $1.7\cdot10^{3}$ \\
 & $256$ & $3.27$ & $3.28$ & $1.00$ & $21.9$ & $4$ & $12$ & $2.7\cdot10^{4}$ \\
 & $4096$ & $3.29$ & $3.30$ & $1.00$ & $85.1$ & $4$ & $12$ & $6.8\cdot10^{6}$ \\
\hline
$e^{2x}$ & $64$ & $3.52$ & $3.53$ & $1.00$ & $12.3$ & $12$ & $16$ & $4.1\cdot10^{3}$ \\
 & $256$ & $3.59$ & $3.61$ & $1.00$ & $24.0$ & $12$ & $16$ & $7.0\cdot10^{4}$ \\
 & $4096$ & $3.61$ & $3.63$ & $1.00$ & $93.5$ & $12$ & $16$ & $1.9\cdot10^{7}$ \\
\hline
$1+9x^{2}$ & $64$ & $3.77$ & $3.79$ & $1.00$ & $13.1$ & $28$ & $28$ & $4.9\cdot10^{3}$ \\
 & $256$ & $3.86$ & $3.87$ & $1.00$ & $25.7$ & $28$ & $28$ & $8.3\cdot10^{4}$ \\
 & $4096$ & $3.88$ & $3.90$ & $1.00$ & $100.4$ & $28$ & $28$ & $2.2\cdot10^{7}$ \\
\hline
$1+\sin(2\pi x)/2$ & $64$ & $3.57$ & $3.59$ & $1.00$ & $12.4$ & $44$ & $44$ & $2.6\cdot10^{3}$ \\
 & $256$ & $3.64$ & $3.66$ & $1.00$ & $24.3$ & $52$ & $52$ & $4.2\cdot10^{4}$ \\
 & $4096$ & $3.66$ & $3.68$ & $1.00$ & $94.6$ & $52$ & $52$ & $1.1\cdot10^{7}$ \\
\hline
\end{tabular}
\caption{The three-point discretization of $-(p(x)w')'$, with $\sigma=0$:
balance $b_{u, v}$ of the generators of $R=(A+i\gamma I)^{-1}$, subnormalization $\beta$
of the Cayley encoding built from them, and degree $d_{\mathrm{sp}}$ of the
polynomial reproducing a generator, defined in the text, at the two widths
$\gamma_1=10^{-6}\Delta$ and $\gamma_2=10^{-1}\Delta$; the last column is the
condition number, which is the order of the degree required by the polynomial
and the exponential approaches.}
\label{tab:jacobi}
\end{table}

Table~\ref{tab:jacobi} shows that both conditions hold on this family. The
balance lies between $3.2$ and $3.9$ for every coefficient, dimension and width,
so that $\beta-1$ grows as $\gamma\sqrt N$, as the two columns of $\beta$ confirm.
The degree $d_{\mathrm{sp}}$ depends on the variation of $p$ and on the width,
from $4$ for the constant coefficient to $52$ for the oscillating one, but not
on $N$, so the state preparations remain polylogarithmic. For the functions
above, when $\eta$ is bounded, the effective cost of the Cayley approach is
therefore $O(b_{u, v} \,\eta\sqrt N\log(1/\varepsilon))=\Ot(\kappa^{1/4})$, against
$\Ot(\kappa)$ for the polynomial and the exponential approaches. Since the
interval $[\Delta,c\Delta]$ lies near the smallest eigenvalue, the polynomial
approach can reduce its degree to $\Ot(\sqrt\kappa)$ by applying QSVT to $2A-I$,
whose spectrum fills $[-1,1]$, which uses the knowledge that $A$ is positive
definite; the Cayley approach remains cheaper, by a factor $\kappa^{1/4}$.

The cost below $\kappa$ does not contradict the $\Omega(\kappa)$ lower bound for
quantum linear system solvers~\cite{CAS22}, which holds when the only access to
$A$ is a block encoding of it. Here $A$ enters through a classical description
of the generators of its resolvent, and the comparison is between access models
rather than between algorithms under a common oracle. The family shows that the
condition of Section~\ref{sec:structured} admits solutions: the cost in gates is
$\Ot(\sqrt N)$, below the cost of the classical computation, provided $b$ is
given as a quantum state and the result is used as one; if $b$ has to be loaded
from a classical vector, or $f(A)b$ read out in full, the cost is $\Omega(N)$.

\paragraph{Code availability.} The implementation is self-contained, requires
no toolboxes, and reproduces every table of this section. The code  is available on Zenodo~\cite{zenodo}.

\section{Conclusions and open problems}\label{sec:open}

We have proposed the Cayley transform as an access model for computing rational functions of a Hermitian matrix on a quantum computer. Used in association with generalized quantum signal processing, it produces the rational functions of type $(2d,2d)$ whose poles lie at $\sigma\pm i\gamma$, each of order at most $d$, with a single ancilla, no
parity restriction, and none of the subnormalization incurred when a rational function is implemented as a linear combination of resolvents. Its two parameters $\sigma$ and $\gamma$ make the encoding tailored to $f$, so that accuracy can be concentrated where it is needed. 
We proved moreover that the products of Cayley factors are exactly the rational functions unitary on every Hermitian matrix, so the choice is forced rather than merely convenient.
Through the change of variable to the circle, the design of the approximant becomes an ordinary trigonometric approximation problem, for which the classical theory is available unchanged.

On the side of cost the encoding admits a unitary realization for every Hermitian $A$. In the query model a fixed-pole rational function is reached exactly at degree one, where a polynomial method pays a degree that grows as the pole approaches the spectrum. Whether this is still an advantage in the realization model depends on how the
encoding is built. We showed that the generic construction from a block
encoding of $A$ cancels the gain (Corollary~\ref{cor:cancellation}), and that
structure in $A$ can bypass that construction, exhibiting one family for which
the shifted solve is obtained from the generators of the
resolvent~\cite{ADV26}.

The first is the cost of loading the data that describe $A$. For a tridiagonal
$A$ the inverse of the shifted matrix is always one-pair semiseparable, but
loading its generators as quantum states at polylogarithmic depth requires a
regularity that holds for the discretized Sturm--Liouville operators of
Section~\ref{sec:numerics-jacobi}, with analytic coefficient, and is not
guaranteed in general. The CMV construction of Section~\ref{sec:cmv} raises the
same question for its coefficients, together with that of characterizing the
matrices whose Cayley transform is a CMV matrix. Beyond the tridiagonal case,
natural candidates for a structured realization are hierarchical matrices,
whose inverses retain a low-rank structure and for which block encodings are
available~\cite{NKL22}, and indefinite operators from partial differential
equations, for which the imaginary part of the shift keeps the resolvent well
conditioned.

The second direction is reaching the full rational class. A product of Cayley
factors realizes the functions $P(b(\lambda))$, whose numerator is tied to the
denominator by unitarity, while the full class with poles at
$\{w_j,\bar w_j\}$, each of order at most $d$, is reached by a linear combination
of $m$ GQSP circuits, at the cost of $\lceil\log_2m\rceil$ further ancillas.
Whether the full class can be reached with fewer ancillas, keeping the
number of queries proportional to the degree, is a question we leave to future
work.

Another direction is a gate-level analysis. Theorem~\ref{thm:endtoend} counts
queries, and concrete gate counts for the structured realization of
Section~\ref{sec:structured}, against QSVT applied to $U_A$ and against QET-U
with a Trotterized exponential, would turn the asymptotic statements into
quantities one can weigh. On structured families one may also process the
block-encoded resolvent directly, which is normal but not Hermitian, through
quantum eigenvalue processing~\cite{LS24}: this avoids the amplification that
makes the encoding unitary, at the price of a polynomial acting on a smaller
spectral region, and the balance between the two remains to be assessed.

\section*{Acknowledgment}
During the preparation of this work the author used Claude (Anthropic) to
assist with drafting and editing the text, running numerical verifications,
and clarifying technical points raised by the author. The author reviewed
and independently verified all results and takes full responsibility for
the content of the paper.
\section*{Funding}
Partial support was provided by the INdAM - GNCS project ``Strutture di matrici e di funzioni per la sintesi di circuiti quantistici efficienti'', CUP E53C23001670001 and by  the U.S. Department of Energy, Office of Science, National Quantum Information Science Research Centers, Superconducting Quantum Materials and Systems Center (SQMS), under Contract No. 89243024CSC000002.
 \bibliographystyle{quantum}
\bibliography{references_clean}
\end{document}